\documentclass[journal]{IEEEtran}  
\usepackage{graphicx}  
\usepackage[caption=false,font=normalsize,labelfont=sf,textfont=sf]{subfig}  
\usepackage{lipsum}  
\usepackage[acronym]{glossaries}  
\usepackage{amsmath}  
\usepackage[font=small]{caption}
\usepackage{url}

\newacronym{das}{DAS}{Distributed acoustic sensing}
\newacronym{dbscan}{DBSCAN}{Density-based spatial clustering of applications with noise}
\newacronym{cnn}{CNNs}{Convolutional neural networks}
\newacronym{ntnu}{NTNU}{Norwegian University of Science and Technology}
\newacronym{cgf}{CGF}{Centre for Geophysical Forecasting}
\newacronym{lpf}{LPF}{Low-pass filter}
\newacronym{iir}{IIR}{Infinite impulse response}
\newacronym{svm}{SVM}{Support vector machine}
\newacronym{snr}{SNR}{Signal-to-noise ratio}
\newacronym{iu}{IU}{Interrogator Unit}
\newacronym{tpe}{TPE}{Tree-structured parzen estimator}
\newacronym{kde}{KDE}{kernel density estimation}
\newacronym{fn}{FN}{false negative}
\newacronym{fp}{FP}{false positive}
\newacronym{tp}{TP}{true positive}
\newacronym{rgb}{RGB}{Red Green Blue}
\newacronym{ssi}{SSI}{spatial sampling interval}
\newacronym{hdf5}{HDF5}{Hierarchical Data Format version 5}

\begin{document}

\title{Edge Computing in Distributed Acoustic Sensing: An Application in Traffic Monitoring}
\author{Khanh Truong, Jo Eidsvik, Robin Andre Rørstadbotnen}
\markboth{Journal, October~2024}%
{Truong \MakeLowercase{\textit{et al.}}: Edge Computing in Distributed Acoustic Sensing: An Application in Traffic Monitoring}
\maketitle

\begin{abstract}
\gls{das} technology leverages fiber optic cables to detect vibrations and acoustic events, which is a promising solution for real-time traffic monitoring. In this paper, we introduce a novel methodology for detecting and tracking vehicles using \gls{das} data, focusing on real-time processing through edge computing. Our approach applies the Hough transform to detect straight-line segments in the spatiotemporal \gls{das} data, corresponding to vehicles crossing the Åstfjord bridge in Norway. These segments are further clustered using the \gls{dbscan} algorithm to consolidate multiple detections of the same vehicle, reducing noise and improving accuracy. The proposed workflow effectively counts vehicles and estimates their speed with only tens of seconds latency, enabling real-time traffic monitoring on the edge. To validate the system, we compare \gls{das} data with simultaneous video footage, achieving high accuracy in vehicle detection, including the distinction between cars and trucks based on signal strength and frequency content. Results show that the system is capable of processing large volumes of data efficiently. We also analyze vehicle speeds and traffic patterns, identifying temporal trends and variations in traffic flow. Real-time deployment on edge devices allows immediate analysis and visualization via cloud-based platforms. In addition to traffic monitoring, the method successfully detected structural responses in the bridge, highlighting its potential use in structural health monitoring.
\end{abstract}



\section{Introduction}

\IEEEPARstart{D}{istributed} acoustic sensing (\gls{das}) is a technology that uses fiber optic cables to detect acoustic signals along the length of the cable. It is based on the principle that a laser light is emitted into a fiber optic cable to sense changes in the cable in response to external forces, causing changes in the backscattered light from impurities in the fiber. By emitting light pulses into the fiber cable one can hence measure the strain along the cable and detect acoustic events of interest. \gls{das} has been widely used in various applications, such as seismic monitoring \cite{fernandez2022seismic}, whale tracking \cite{rorstadbotnen2023simultaneous}, pipeline monitoring \cite{muggleton2020gas}, ship detection \cite{landro2022sensing} and structural health monitoring \cite{hubbard2021dynamic}.

In recent years, various methods on spatiotemporal \gls{das} data have been proposed for traffic monitoring, road maintenance, and safety purposes. Accordingly, the vehicle presence can be detected by applying a dual-threshold algorithm to analyze the energy and zero-crossing rates of the signals, achieving accuracy above 80\% \cite{liu2019vehicle}. The according speed can be estimated within an 5\% of error by calculating the time it takes for a vehicle to pass through multiple detection points along the optical fiber. Then supervised machine learning like \gls{svm} can be used to classify to either cars, SUVs or trucks based on the features extracted from denoised signals \cite{liu2019vehicle}. Alternative solutions rely on linear regression. For identifying the start and end points of each vehicle’s passage along the fiber \cite{chiang2023distributed}. Along with labeled data, one can train a \gls{cnn} to detect the vehicle type and size, with accuracy reaching up to 94\% for classifying vehicle types and 95\% for classifying vehicle sizes. Also employing deep neural network, \cite{wang2021research} focused on identifying high-speed railway's events such as track cracking, beam crevices and switches. Their model uses data augmentation by combining vibration data collected at different times to form three-channel data in an \gls{rgb} color format. The overall accuracy from various \gls{cnn} architectures reaches 98\% (VGG-16, ResNet). Pre-classifier events effectively reduces the \acrlong{fn} rate by approximately 60\%.

A more recent approach to detect vehicles is the Hough transform, which can identify straight lines in the data \cite{thomas2023performance, wiesmeyr2021distributed}. The methods accurately estimated the number of trucks on a highway in Austria, with a median difference of one vehicle per one-minute interval (52\% of the intervals showing no difference), as well as the average speed of trucks (deviations of ±10 km/h) \cite{wiesmeyr2021distributed}. The method was also applied to \gls{das} data along a highway in Norway, where signal qualities such as \gls{snr} and continuity have been used for evaluation \cite{thomas2023performance}.

Despite these promising results, several challenges remain to be solved. The traditional Hough transform is highly sensitive to parameter settings; an excessively high threshold can miss smaller vehicles with weaker signals \cite{wiesmeyr2021distributed}, while a low threshold may yield numerous false positives. Furthermore, the literature has yet to address real-time deployment and edge computing for such approaches. Considering the massive data sizes of \gls{das}, this is not straightforward because signal processing routines and statistical analysis can be computational expensive. 

In the current paper, we introduce a novel methodology to detect and track vehicles in real-time using \gls{das} data. The main contribution is a workflow of steps for fast detection and estimation of traffic events. The proposed method is based on the Hough transform to detect straight line segments in the \gls{das} data, followed by density-based spatial clustering of applications with noise (\gls{dbscan}) to cluster duplicate line segments. The methodology is validated on simultaneous \gls{das} and camera data from the Åstfjord bridge, Norway (Figure \ref{fig:trondheim}). Results are very promising, delivering counts and velocities with a high level of accuracy and efficiency.  Results show that we can accurately count the number of cars and trucks crossing the bridge in both directions, as well as estimate their velocity. Importantly, the workflow enables real-time data processing and statistical analysis that is computed on the edge and visualized on the fly.

In Section \ref{sec:das}, we provide background on \gls{das} data and the data set gathered at the Åstfjord bridge. In Section \ref{sec:methods}, we outline our methods for \gls{das} data processing, statistical analysis and real-time deployment of the  detection and estimation routines. In Section \ref{sec:results}, we present results of the methodologies and illustrate various outputs that are relevant for traffic monitoring. In Section \ref{sec:conclusion}, we conclude and point to interesting future work.

\section{\gls{das} and Åstfjord bridge data}
\label{sec:das}

\subsection{\gls{das}}

\gls{das} is a technology that utilizes fiber optic cables to gauge acoustic signals along the length of the cable. The overall technical process of \gls{das} starts with a pulse of light that is transmitted down the fiber optic cable from an \gls{iu}. As this pulse travels along the cable, it comes across impurities within the fiber, causing scattering of the light. The portion of the light that is scattered back to the light source is referred to as Rayleigh backscattering. When an acoustic wave, such as a vibration or sound, interacts with the fiber, it strains the fiber at the locations influenced by this acoustic wave. These changes in the fiber's length cause a corresponding modification in the backscattered signal. 

\begin{figure}[htbp]
    \centering
    \includegraphics[width=0.30\textwidth]{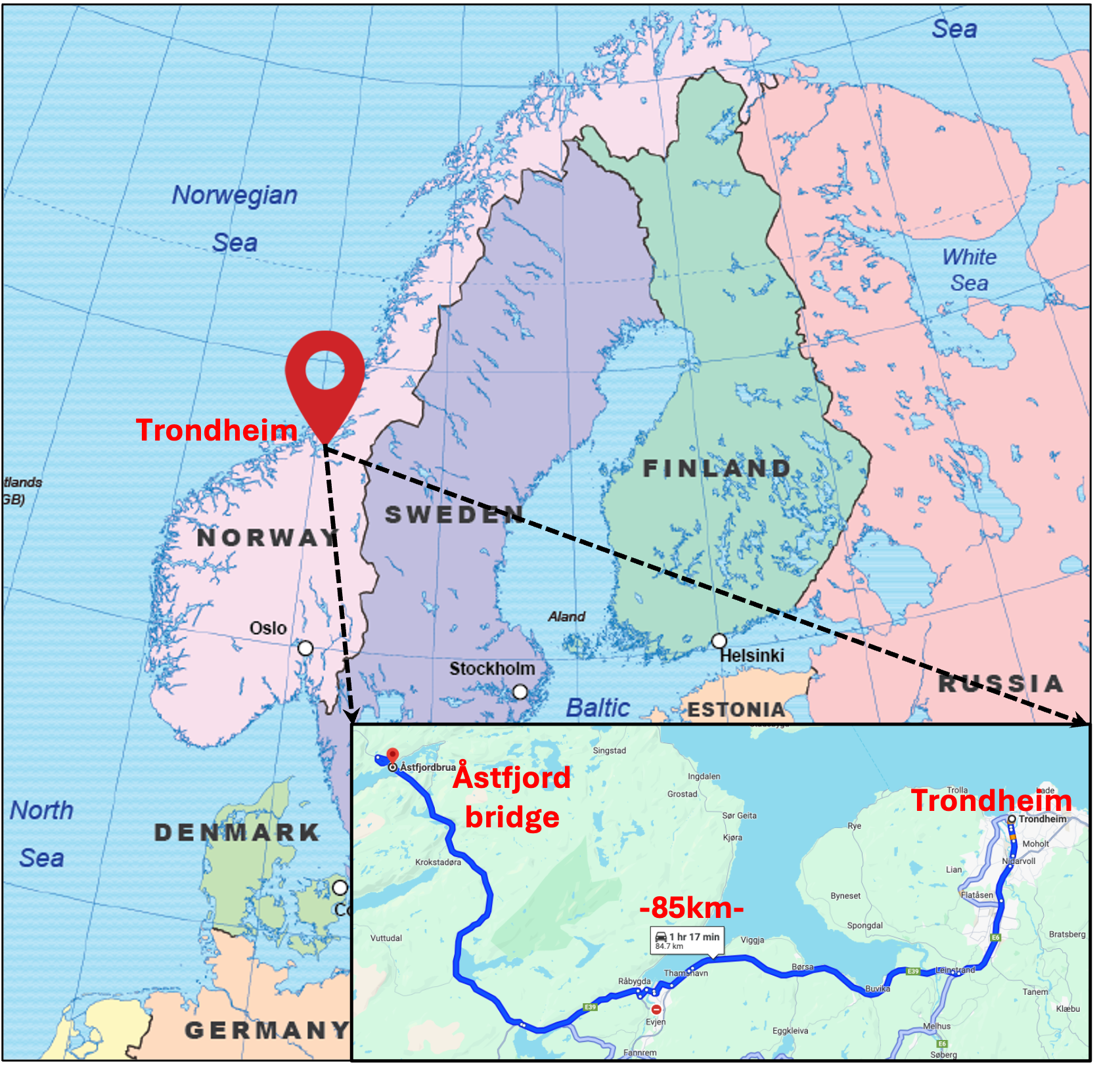}
    \caption{The Åstfjord bridge is located 85 km from Trondheim, Norway.}
    \label{fig:trondheim}
\end{figure}

Specially, the phase $\varphi(s)$ of the backscattered light at position $s$ is given by $\varphi(s) = \beta \cdot \epsilon(s)$ where $\beta$ is a constant related to the optical properties of the fiber, and $\epsilon(s)$ is the strain at this position. The \gls{iu} measures the phase changes at discrete points along the fiber, known as channels. The \gls{ssi} is determined by the pulse repetition rate and the speed of light in the fiber. Given the sampling period $\Delta\tau$ and speed of light in vacuum $c$, the SSI is $\frac{\Delta\tau \cdot c}{2n_g}$ where $n_g$ is the group refractive index of the fiber. The strain rate $\dot{\epsilon}(s,t)$ at location $s$ and time $t$ is derived from the time-differentiated phase change $\dot{\epsilon}(s,t) = \frac{\partial \varphi(s,t)}{\partial t}$. By analyzing this time-differentiated phase change over space and time, \gls{das} can identify and locate acoustic events or disturbances that occur near the cable \cite{taweesintananon2021distributed}.

\subsection{Åstfjord bridge data}

\begin{figure}[htbp]
    \centering
    \subfloat[]{\includegraphics[trim={4in 0in 4in 1.43in}, clip, width=0.23\textwidth]{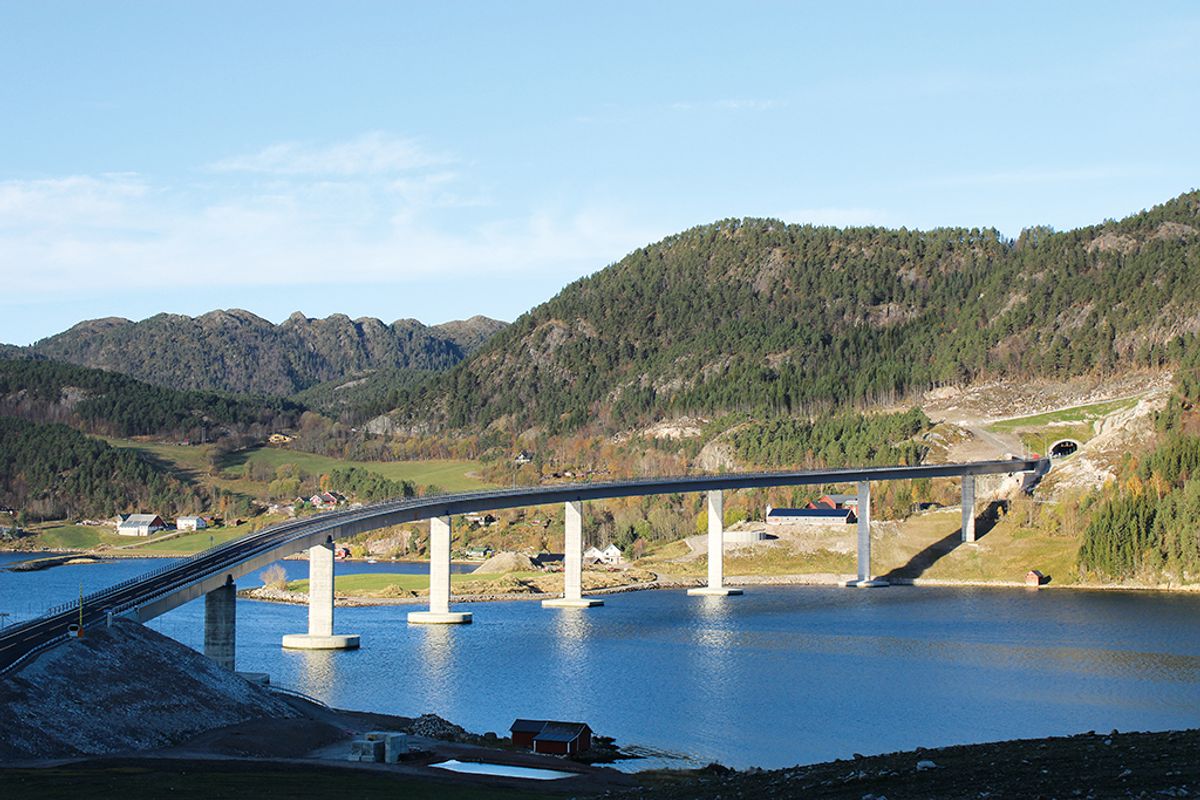}}
    \label{fig:bridge-overview}
    \hfill
    \subfloat[]{\includegraphics[trim={0in 3in 0in 2in}, clip, width=0.25\textwidth]{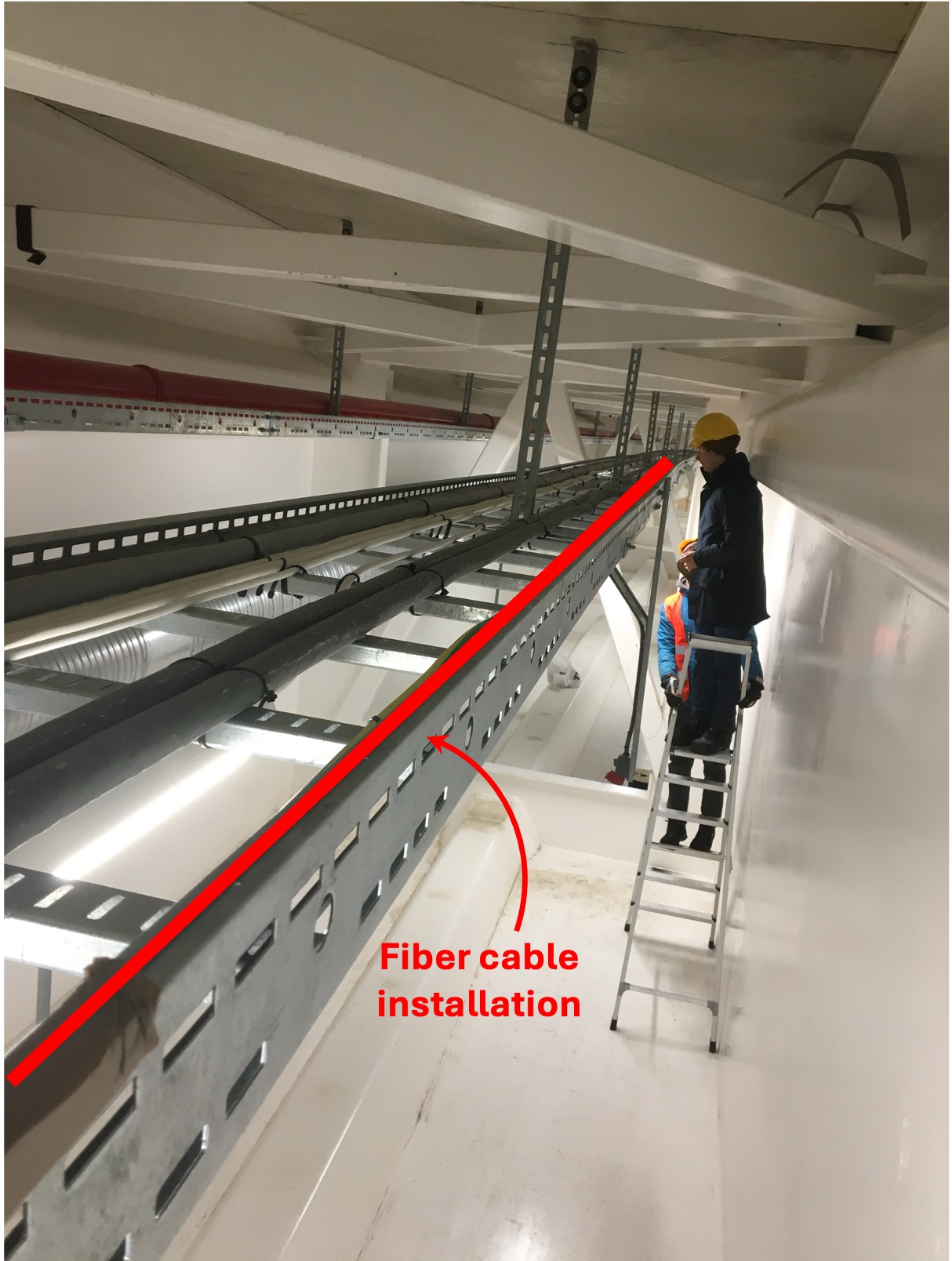}}
    \label{fig:cable-install}
    \caption{(a) The Åstfjord bridge spans 735 meters. (b) The fiber cable installation in the inspection walkway under the bridge surface was done by \gls{cgf} in February 2023.}
    \label{fig:bridge}
\end{figure}

Åstfjord is approximately 85 km from Trondheim, Norway. In 2017, the government initiated the construction of a bridge across the fjord (Figure \ref{fig:trondheim}). The Åstfjord bridge is 735 meters long, comprising eight spans, with the longest being 100 meters. It was completed and opened for traffic in February 2021. In February 2023, in collaboration with the county council, the \gls{cgf} (\url{www.ntnu.no/cgf}) installed a fiber cable in the inspection walkway and connected it to an \gls{iu} (Figure \ref{fig:bridge}). Since then, \gls{das} data have been continuously recorded and stored on-site, capturing vibrations caused by for instance vehicle crossings and wind.

The data features a spatial resolution of 1 meter and a temporal resolution of 1000 Hz. It is stored in \gls{hdf5} format, organized in 10-second batches. Each  \gls{hdf5} file is approximately 16 MB, resulting in a daily data volume of roughly 140 GB.

\begin{figure}[htbp]
    \centering
    \includegraphics[width=0.35\textwidth, trim={0in, 0in 0in 0in}, clip]{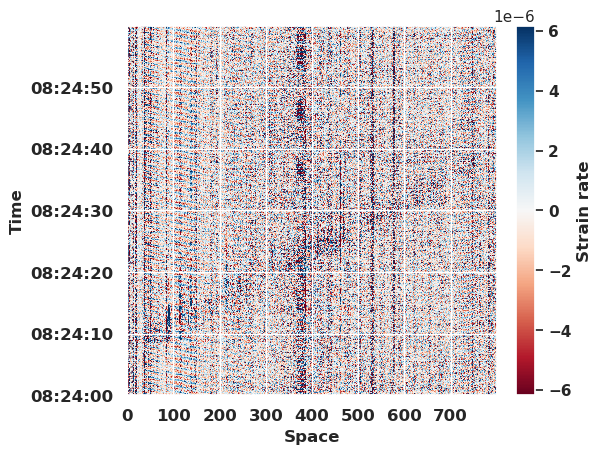}
    \caption{A 60-second sample of \gls{das} data (strain rate) from the Åstfjord bridge in the morning on 5 October 2023. The X-axis represents channel index, and the Y-axis represents time in UTC. Coherent strain rate data along a straight line indicate a vehicle crossings during this period (08:24:09 - 08:24:40).}
    \label{fig:strain-rate}
\end{figure}

In addition to the \gls{das} data, we recorded traffic data by camera at two ends of the bridge for validation purposes. This was done for a few hours in October 2023. Based on the camera information, we know the true times of vehicle crossing and the type of vehicles crossing the bridge during this time period. This ground truth is highly valuable for testing our suggested methodology for detection and estimation of events in the \gls{das} data.

\section{Methodology}
\label{sec:methods}

Because the cable is installed along the bridge, vehicles crossing it at constant velocity generate straight line signals in the \gls{das} data. However, the signal coupling from bridge surface to the fiber is not straightforward and there is noise caused by wind and other external sources. Hence, coherent line signals are not always easy to discern in the data, especially for small vehicles. In Figure \ref{fig:strain-rate}, we show a typical \gls{das} strain rate display. Here, a vehicle can be seen to enter the bridge at time 08:24:09 and drive off at 08:24:40. To automate the detection of such lines and their signal characteristics, we suggest a workflow of several processing and analysis steps. The methodology starts with some preprocessing steps, primarily to clean, down-sample and threshold the data. Afterwards, the Hough transform is used to detect straight line segments in the image. This will output the coordinates of detected line segments. Some of them may be very close to each other due to noise. Hence, \gls{dbscan} is employed to group close lines. We will now describe all of these steps.

\subsection{Data preprocessing}

When placing the fiber cable over the bridge, there are some redundant parts that do not contribute to the seismic recording. The first 36 channels and the last 49 channels are spare near the \gls{iu} and the ending spare, respectively. There is another 22-channel length fiber coil (from channel 365 to 386) near a door in the walkway. Overall, there is a final 693 channels length of \gls{das} data that we use for traffic monitoring. This \gls{das} strain rate data goes through five preprocessing steps:
\begin{enumerate}
    \item Low-pass filtering: remove noise,
    \item Down-sampling: reduce temporal sampling of the data,
    \item Gaussian smoothing: enhance signal in driving directions,
    \item Sobel filtering: highlight sharp transitions in the data,
    \item Binary thresholding: keep relevant transition pixels, 
\end{enumerate}
We next describe each of these in more detail.

\subsubsection{Low-pass filtering}

A \gls{lpf} allows signals with frequency $f$ lower than a specific cutoff frequency to pass while attenuating higher frequencies. The frequency response $ H(f) $ of an ideal \gls{lpf} is defined as:
\begin{equation}
    H(f) = \begin{cases}
        1 & |f| \leq f_c, \\
        0 & |f| > f_c,
    \end{cases}
\end{equation}
where $ f_c $ is the cutoff frequency.

\begin{figure}[htbp]
    \centering
    \includegraphics[width=0.35\textwidth, trim={0in, 0in 0in 0in}, clip]{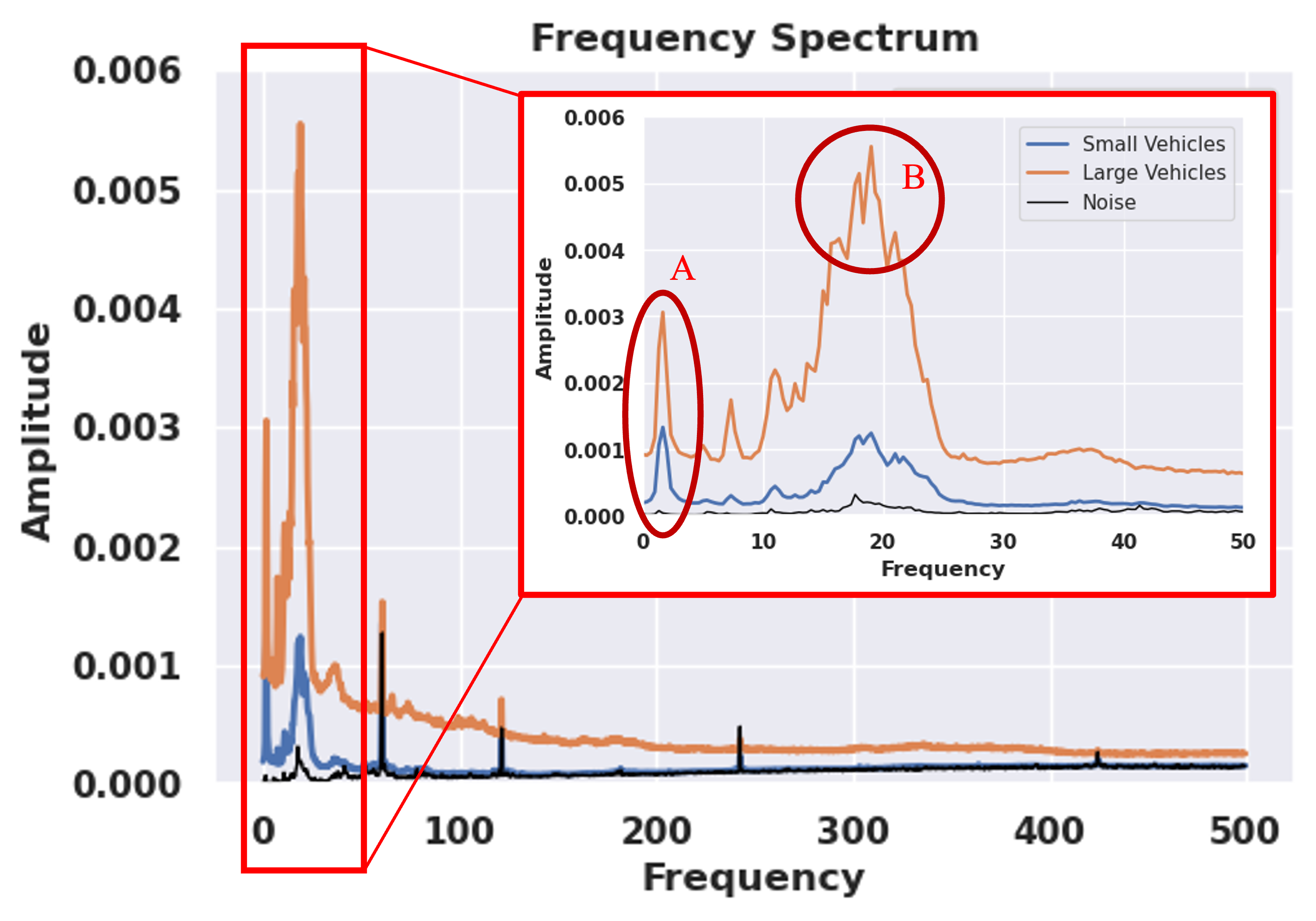}
    \caption{The fast Fourier transform converts time domain to frequency domain. The orange line represents the frequency spectrum of a large vehicle (truck). The blue line is that of a the small vehicle. The black line represents background noise. Part A of the spectrum is the quasi-static deformation signals ($<$1Hz) and part B is the vehicle-induced surface waves (15 - 25 Hz) \cite{yuan2021urban, liu2024characterizing}. We focus on the low frequency A for identifying car movement.}
    \label{fig:fft}
\end{figure}

\begin{figure}[htbp]
    \centering
    \subfloat[]{\includegraphics[trim={0in 0.5in 0.3in 0in}, clip, width=0.255\textwidth]{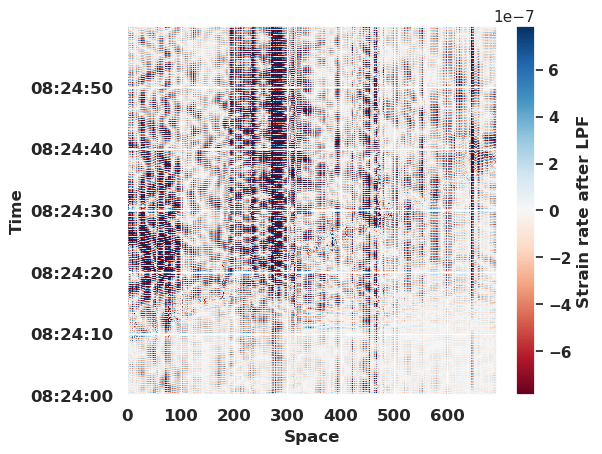}\label{fig:lpf-2.0}}
    \hfill
    \subfloat[]{\includegraphics[trim={1.2in 0.5in 0in 0in}, clip, width=0.225\textwidth]{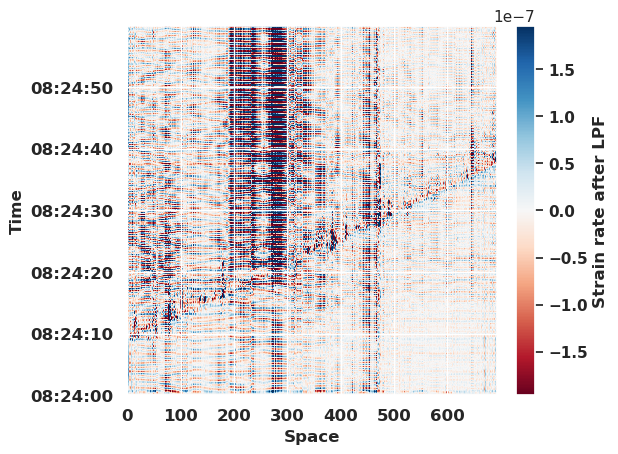}\label{fig:lpf-1.0}}
    \hfill
    \subfloat[]{\includegraphics[trim={0in 0in 0.3in 0in}, clip, width=0.26\textwidth]{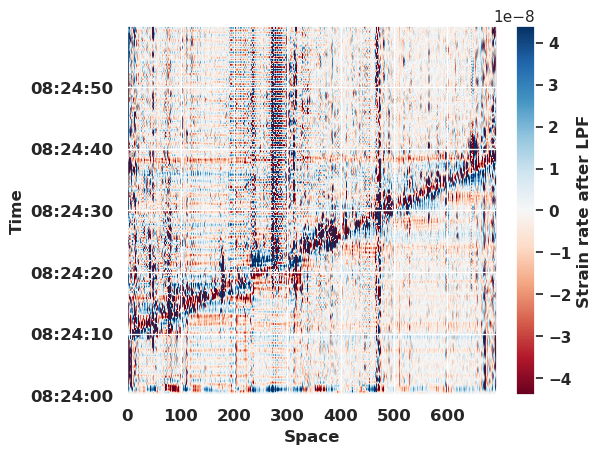}\label{fig:lpf-0.5}}
    \hfill
    \subfloat[]{\includegraphics[trim={1.2in 0in 0in 0in}, clip, width=0.22\textwidth]{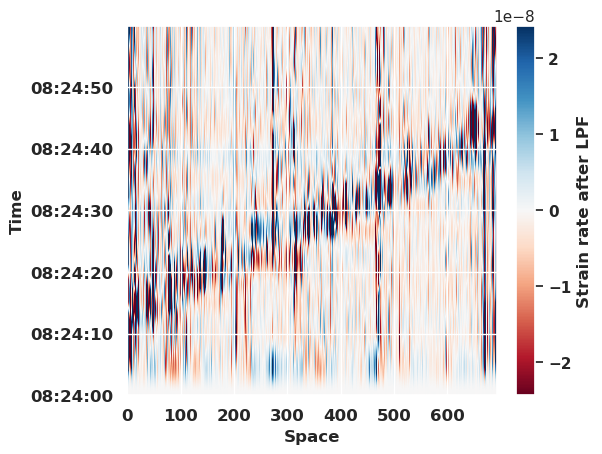}\label{fig:lpf-0.1}}
    \caption{\gls{lpf} with various cutoff thresholds: (a) $f_c=2$, (b) $f_c=1$, (c) $f_c=0.5$, and (d) $f_c=0.1$. With $f_c=2$ the data still contains unnecessary high frequency content. With $f_c=0.1$, it starts blurring out the car's position. The optimal cutoff frequency is fine-tuned using a loss function together with an optimization algorithm.}
    \label{fig:lpf}
\end{figure}

When vehicles drive along the bridge, they generate two types of signals: quasi-static deformation signals (below 1-2 Hz) resulting from their weight and vehicle-induced surface waves (2 - 30 Hz) resulting from the vehicle-road interactive dynamics \cite{yuan2021urban, liu2024characterizing}. Figure \ref{fig:fft} compares the frequency content of 2-second windows with two crossing vehicles and when there are no vehicles crossing, the background noise. There are two obvious peaks in the frequency content of the car's signal, one around 1 Hz and the other around 20 Hz. The background noise, on the other hand, has more uniform frequency content with some harmonic spikes due to an industrial fan close to the \gls{iu}. Given our objective of identifying the position of vehicles, we applied a \gls{lpf} to retain only the low-frequency content. Figure \ref{fig:lpf} shows the data after applying \gls{lpf} with different cutoff frequencies. The optimal cutoff frequency is fine-tuned later using a loss function together with an optimization algorithm.

\subsubsection{Down-sampling}

The original data has a temporal resolution of 1000~Hz, which is more than necessary for vehicle tracking. Therefore, we down-sampled the data to reduce computational complexity while retaining sufficient information about vehicles. The typical speed of vehicles crossing the bridge is around $ v = 85$~km/h. Given the cable length inside the bridge is 693~m, the average time for a vehicle to cross the bridge is $ t = 29.35$~s. To maintain a speed resolution of 0.5 km/h around the typical speed, i.e., distinguishing between speeds of $v = 85$~km/h and $v' = 85.5$~km/h ($t' = 29.18$~s), we need to maintain the temporal sampling interval at maximum $\lvert t - t' \rvert = 0.17 $~s. Thus, the temporal sampling rate must be at least $ \frac{1}{\lvert t - t' \rvert} = 6$~Hz to maintain this level of speed resolution. We choose to down-sample the data to 8~Hz, which is the smallest divisor of the original 1000~Hz that is greater than 6~Hz.

\subsubsection{Gaussian smoothing}

\begin{figure}[htbp]
    \centering
    \includegraphics[width=0.6\linewidth]{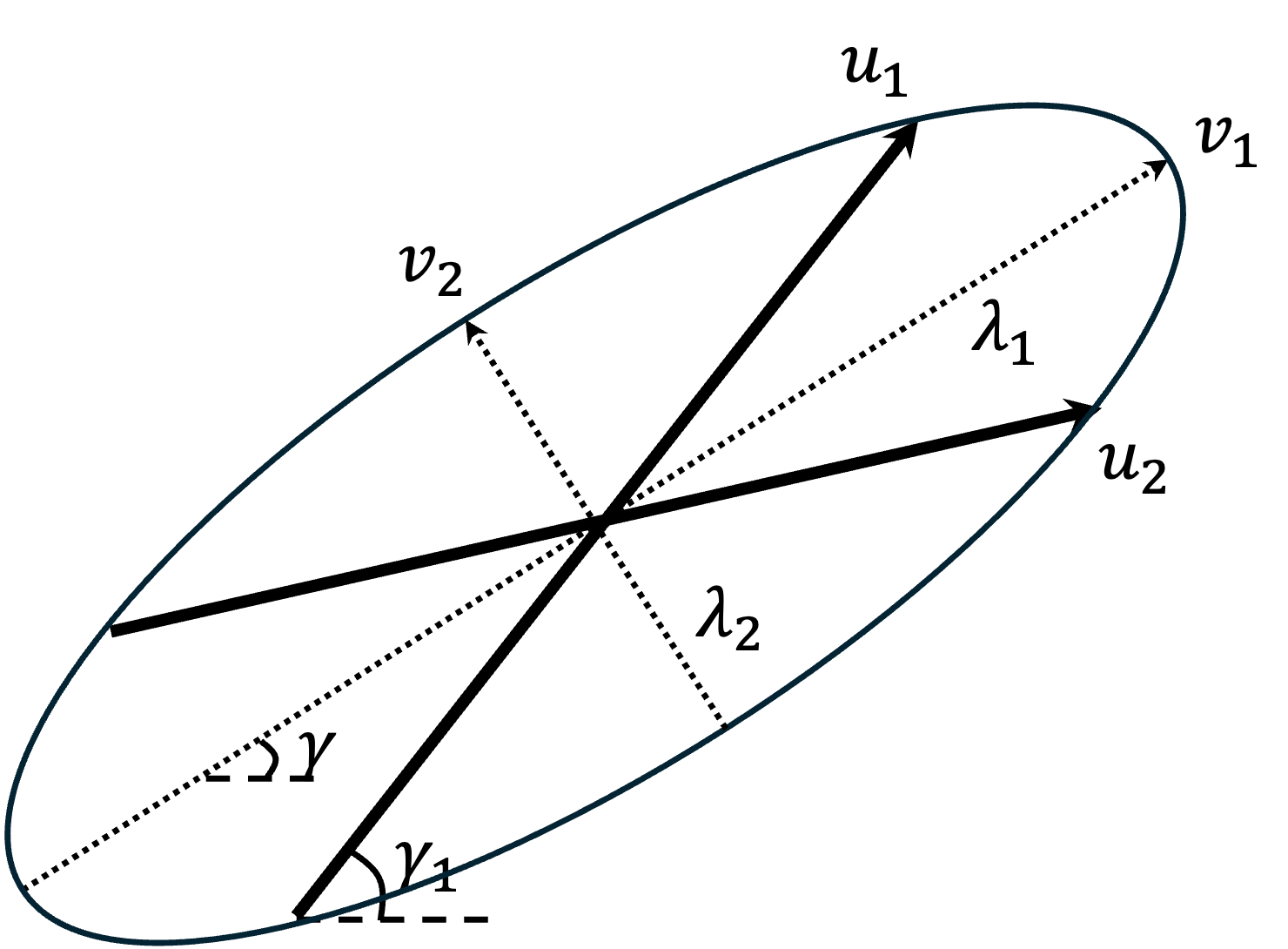}
    \caption{The eigenvalues and eigenvectors of the covariance matrix $\Sigma$. The first eigenvector $v_1$ is in the direction of the desired signal (the average speed of $u_1$ and $u_2$), while the second eigenvector $v_2$ is orthogonal to $v_1$. The eigenvalues $\lambda_1$ and $\lambda_2$ determine the spread of the kernel.}
    \label{fig:eigval-eigvec}
\end{figure}

Gaussian smoothing is a technique using a linear filter that convolves the input signal with a 2D Gaussian kernel \cite{hsiao2007generic}. The formula for a 2D Gaussian kernel is:
\begin{equation}
    f(\mathbf{x}) = \frac{1}{(2\pi |\Sigma|)^{1/2}} \exp\left(-\frac{1}{2} \mathbf{x}^\top \Sigma^{-1} \mathbf{x}\right),
\end{equation}
where $ \mathbf{x} = (s, t)^\top $ represents space-time vector, and $ \Sigma$ is the 2x2 covariance matrix:
\begin{equation}
    \Sigma = \begin{pmatrix} \sigma_s^2 & \sigma_{st} \\ \sigma_{st} & \sigma_t^2 \end{pmatrix}.
\end{equation}
Designing the covariance matrix $\Sigma$ is crucial for capturing directional signals from vehicle movement. The matrix is determined based on the expected speed range, $(u_1, u_2)$, and the standard deviation in spatial dimension, $\sigma_s$. To do this, we decompose $\Sigma$ as:
\begin{equation}
    \Sigma = V \Lambda V^T, \label{eq:eigendecompose}
\end{equation}
where $V$ is the eigenvector matrix and $\Lambda$ is the eigenvalue matrix:
\begin{equation}
    V
    = \begin{pmatrix} v_1 & v_2 \end{pmatrix}
    = \begin{pmatrix} v_{11} & v_{21} \\ v_{12} & v_{22} \end{pmatrix}, \hspace{8mm}
    \Lambda = \begin{pmatrix} \lambda_1 & 0 \\ 0 & \lambda_2 \end{pmatrix}.
\end{equation}
Performing matrix multiplication and denoting $k= \frac{\lambda_1}{\lambda_2}$, equation (\ref{eq:eigendecompose}) can be rewritten as:
\begin{equation}
    \Sigma = \sigma^2_{s}\begin{pmatrix} 1 & \frac{ v_{11} v_{12} ( k -1 )}{k v_{11}^2 + v_{12}^2} \\ \frac{v_{11} v_{12} ( k -1 )}{k v_{11}^2 + v_{12}^2} & \frac{ k v_{12}^2 + v_{11}^2 }{k v_{11}^2 + v_{12}^2} \end{pmatrix}.
\end{equation}
Let $\gamma_1$ and $\gamma$ are the angles corresponding to $u_1$, and the average of $u_1$ and $u_2$, respectively (Figure \ref{fig:eigval-eigvec}). The ratio of eigenvalues can be derived as $k = \frac{1}{\tan(\gamma_1 - \gamma)}$. The eigenvector matrix $V$ becomes:
\begin{equation}
V = \begin{pmatrix} 1 & \tan(\gamma) \\ \tan(\gamma) & -1 \end{pmatrix}.
\end{equation}
Thus, $\Sigma$ is uniquely determined by the vehicle speed range $(u_1, u_2)$ and the spatial standard deviation $\sigma_s$. The kernel is then convolved with the data to enhance directional signals from the moving vehicles.

\begin{figure}[htbp]
    \centering
    \subfloat[]{\includegraphics[trim={0in 0in 1.5in 0in}, clip, width=0.15\textwidth]{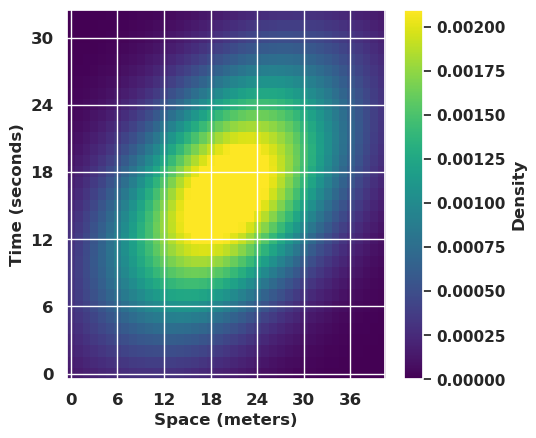}\label{fig:gauss-kernel-wide}}
    \hfill
    \subfloat[]{\includegraphics[trim={0.3in 0in 1.5in 0in}, clip, width=0.14\textwidth]{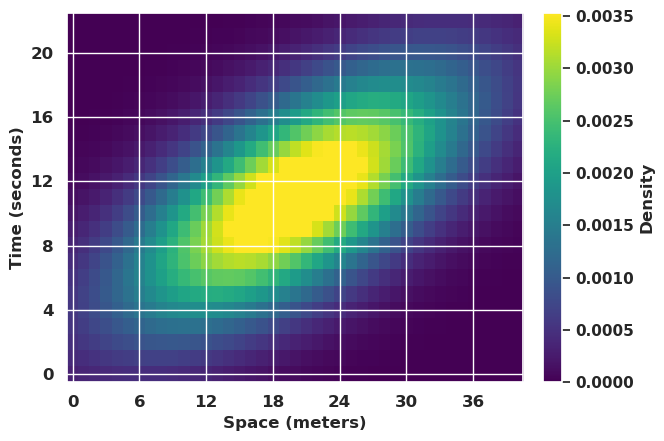}\label{fig:gauss-kernel-normal}}
    \hfill
    \subfloat[]{\includegraphics[trim={0.3in 0in 0in 0in}, clip, width=0.16\textwidth]{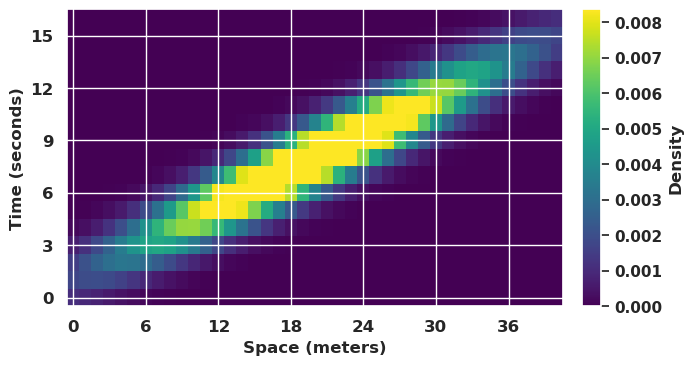}\label{fig:gauss-kernel-narrow}}
    \hfill
    \subfloat[]{\raisebox{0.2cm}{\includegraphics[trim={0in 0.18in 1.7in 0in}, clip, width=0.168\textwidth]{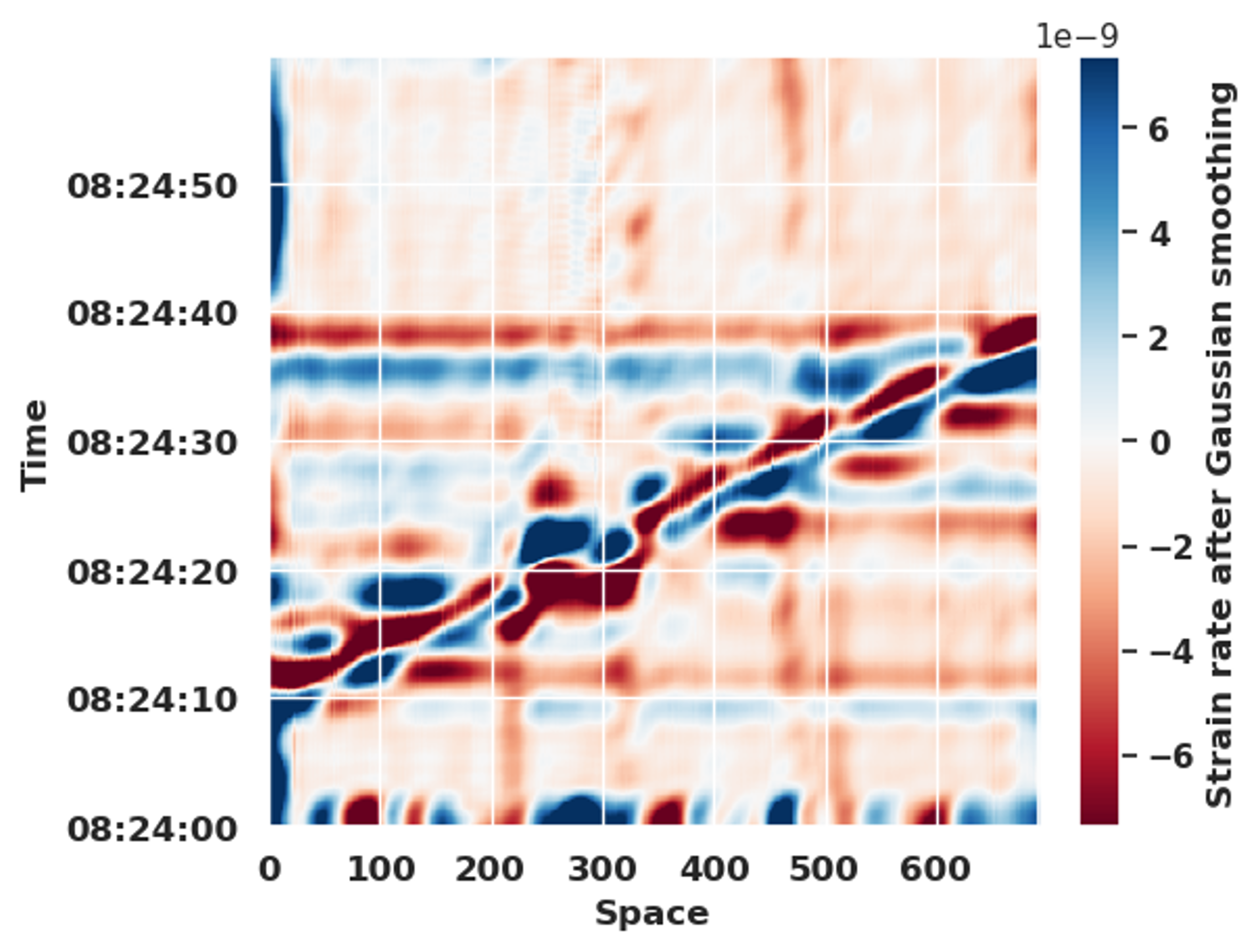}}\label{fig:gauss-smooth-wide}}
    \hfill
    \subfloat[]{\raisebox{0.12cm}{\includegraphics[trim={1.7in 0.01in 1.7in 0in}, clip, width=0.13\textwidth]{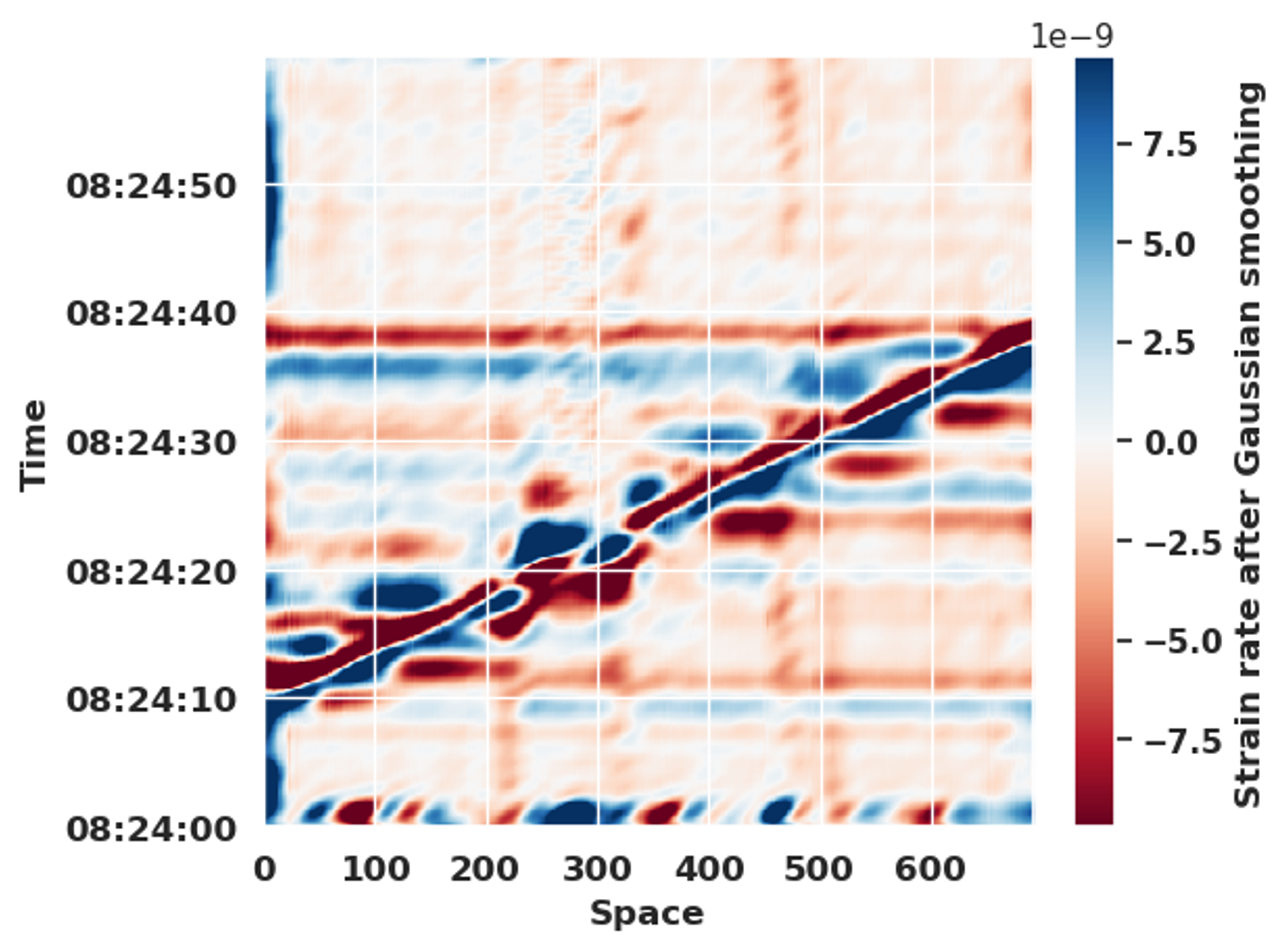}}\label{fig:gauss-smooth-normal}}
    \hfill
    \subfloat[]{\raisebox{0.1cm}{\includegraphics[trim={1.7in 0in 0in 0in}, clip, width=0.175\textwidth]{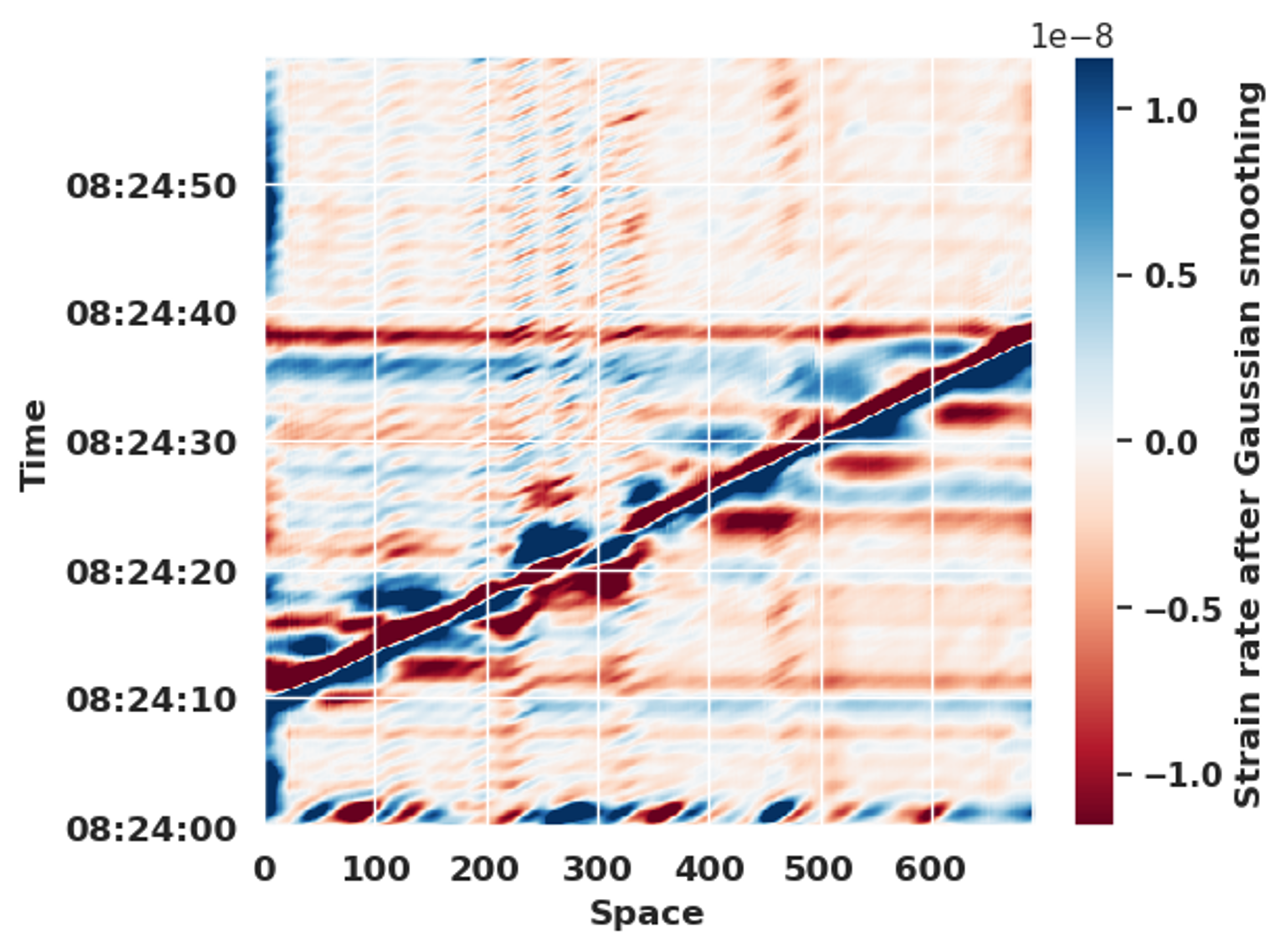}}\label{fig:gauss-smooth-narrow}}
    \caption{Gaussian smoothing with different velocity ranges (with the same $\sigma_{s}=10 $ meters). (a) The kernel with wide speed range (70-100 km/h). (b) The kernel with medium speed range (80-90 km/h). (c) The kernel with narrow speed range (85-86 km/h). The corresponding smoothed data are shown in (d), (e), and (f).}
    \label{fig:gauss-smooth}
\end{figure}

Figure \ref{fig:gauss-smooth} compares the results of Gaussian smoothing with different speed ranges. As depicted, large speed range kernels (Figure \ref{fig:gauss-kernel-wide}, \ref{fig:gauss-smooth-wide}) do not emphasize the straight line signals as clearly as the small speed range kernel (Figure \ref{fig:gauss-kernel-narrow}, \ref{fig:gauss-smooth-narrow}). On the other hand, a narrow speed range may miss signals outside its limits. We choose the medium speed range, between $u_1=80$~km/h and $u_1=90$~km/h (Figure \ref{fig:gauss-kernel-normal}, \ref{fig:gauss-smooth-normal}), to balance these factors.

\subsubsection{Sobel filtering}

The Sobel operator is a discrete differentiation operator used to compute an approximation of the gradient of the image intensity function \cite{kanopoulos1988design}. It uses two convolution kernels (one for the spatial $s$ direction and one for the temporal $t$ direction) to compute the gradients. The gradients are typically large at edge characteristics in the image. Sobel kernels are defined as follows:
\begin{equation}
    G_s = \begin{pmatrix} -1 & 0 & 1 \\ -2 & 0 & 2 \\ -1 & 0 & 1 \end{pmatrix} \text{ and }
    G_t = \begin{pmatrix} -1 & -2 & -1 \\ 0 & 0 & 0 \\ 1 & 2 & 1 \end{pmatrix} \text{.}
\end{equation}
Convolution is applied to the spatiotemporal image with these kernels. Letting \( I \) be the input image, the gradient images in the spatial and temporal directions are obtained by
\begin{equation}
    \hat{I_s} = G_s * I, \hspace{8mm} \hat{I_t} = G_t * I.
\end{equation}

As depicted in Figure \ref{fig:gauss-smooth}, there is a transition from negative to positive values at the location of the crossing vehicle. To enhance this positive gradient, we apply a rectification step where only positive gradients are retained. This means that any negative values in the gradient components are set to zero;
\begin{equation}
    I_s = \max(\hat{I_s}, 0), \hspace{8mm}
    I_t = \max(\hat{I_t}, 0).
\end{equation}

Then the gradient magnitude is calculated by:
\begin{equation}
    |\Delta I| = \sqrt{I_s^2 + I_t^2}.
\end{equation}

\begin{figure}[htbp]
    \centering
    \subfloat[]{\includegraphics[trim={0in 0in 1.52in 0in}, clip, width=0.243\textwidth]{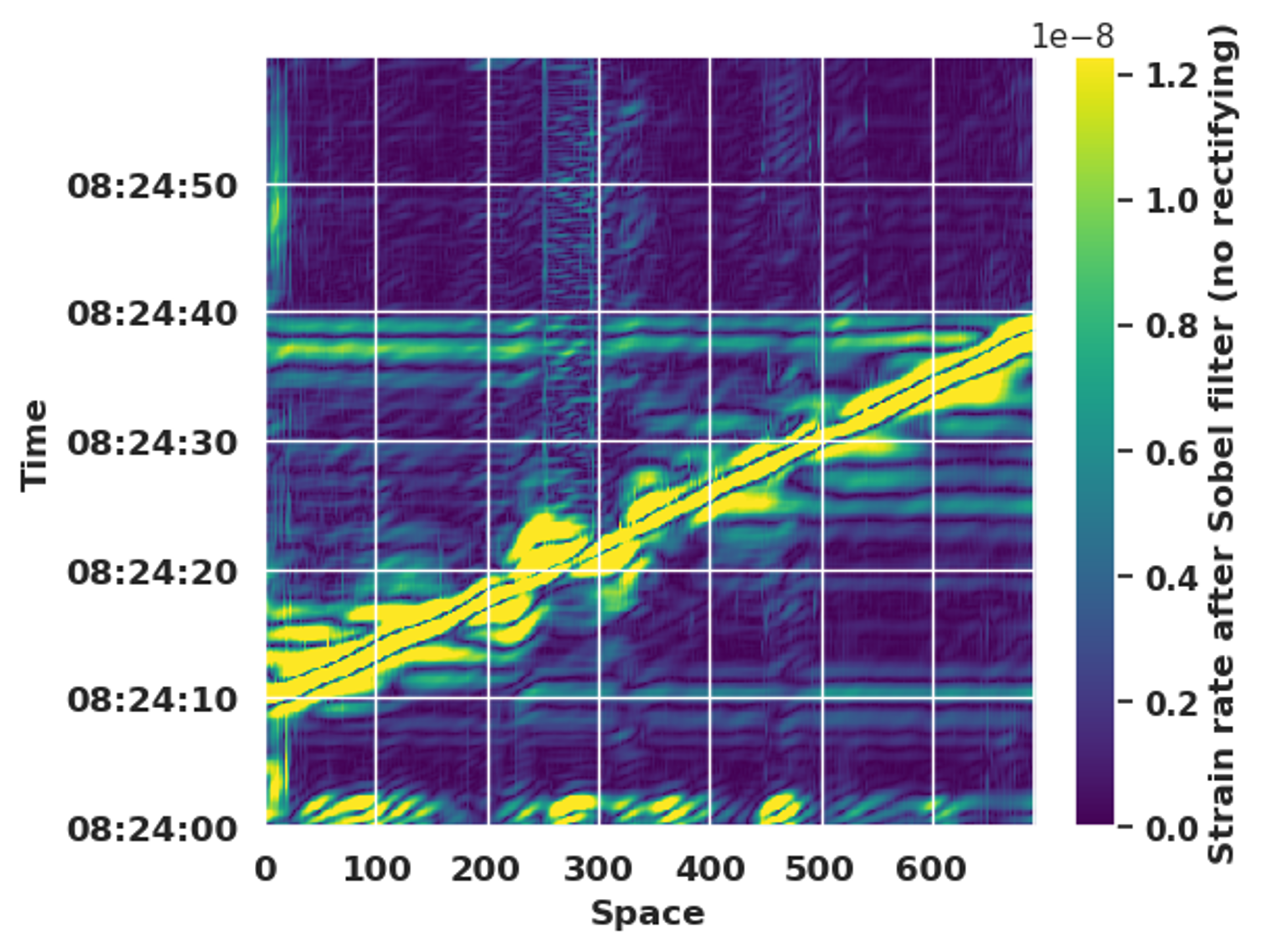}\label{fig:sobel-all}}
    \hfill
    \subfloat[]{\includegraphics[trim={1.6in 0in 0in 0in}, clip, width=0.24\textwidth]{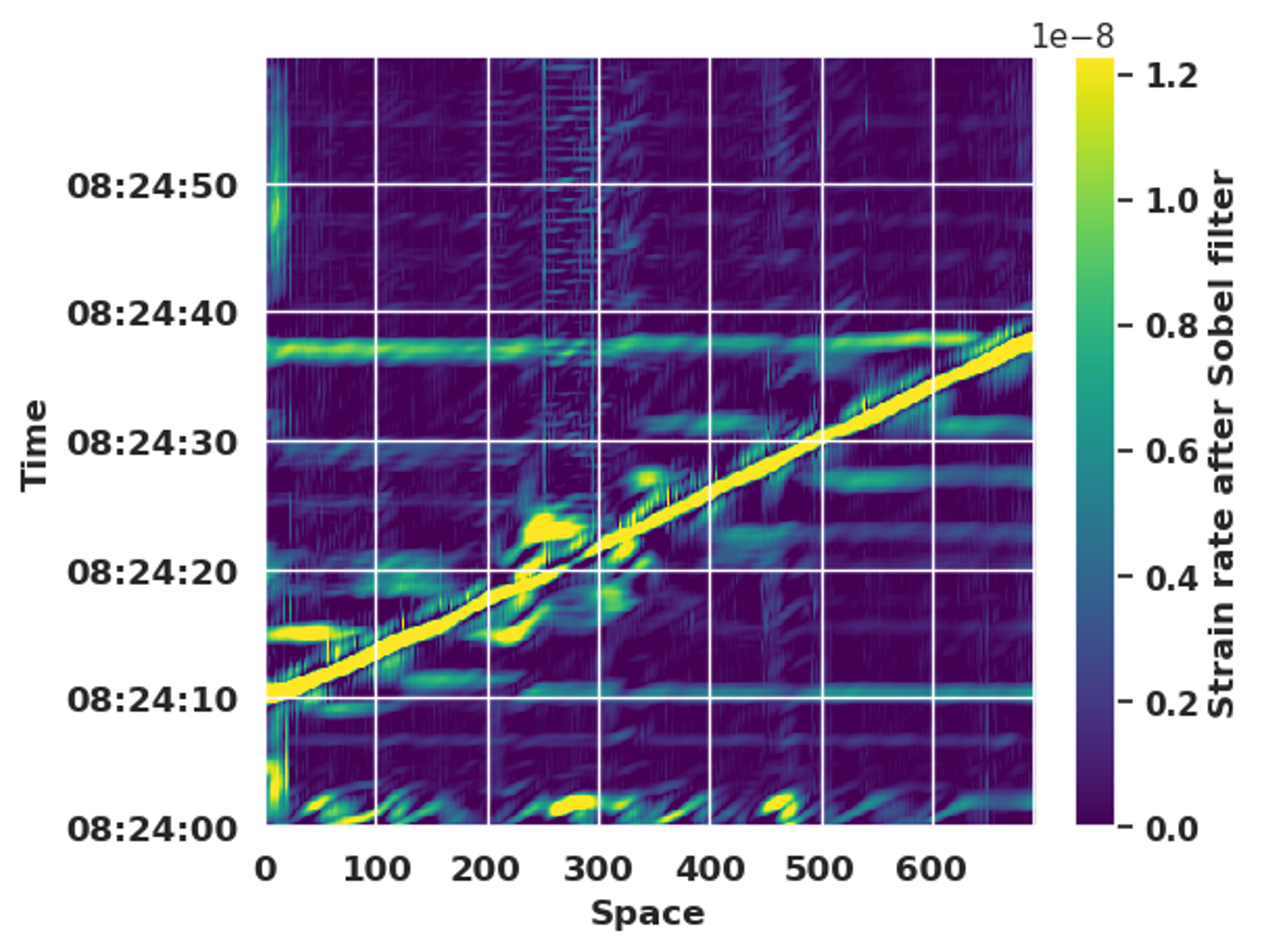}\label{fig:sobel-pos}}
    \caption{Sobel filtering. (a) Gradient magnitude in image without rectifying the negative values. (b) Gradient magnitude in image rectifying the negative values.}
    \label{fig:sobel}
\end{figure}

Figure \ref{fig:sobel} shows the gradient magnitude of the data after applying the Sobel filter. The gradient magnitude image highlights the sharp transitions in the data, which are indicative of vehicle crossings. Rectifying the negative values (Figure \ref{fig:sobel-pos}) gives a significant improvement in the preservation of desired pixels, compared to values are not being rectified (Figure \ref{fig:sobel-all}).

\subsubsection{Binary thresholding}

\begin{figure}[htbp]
    \centering
    \subfloat[]{\includegraphics[trim={0.3in 0in 1.2in 0in}, clip, width=0.171\textwidth]{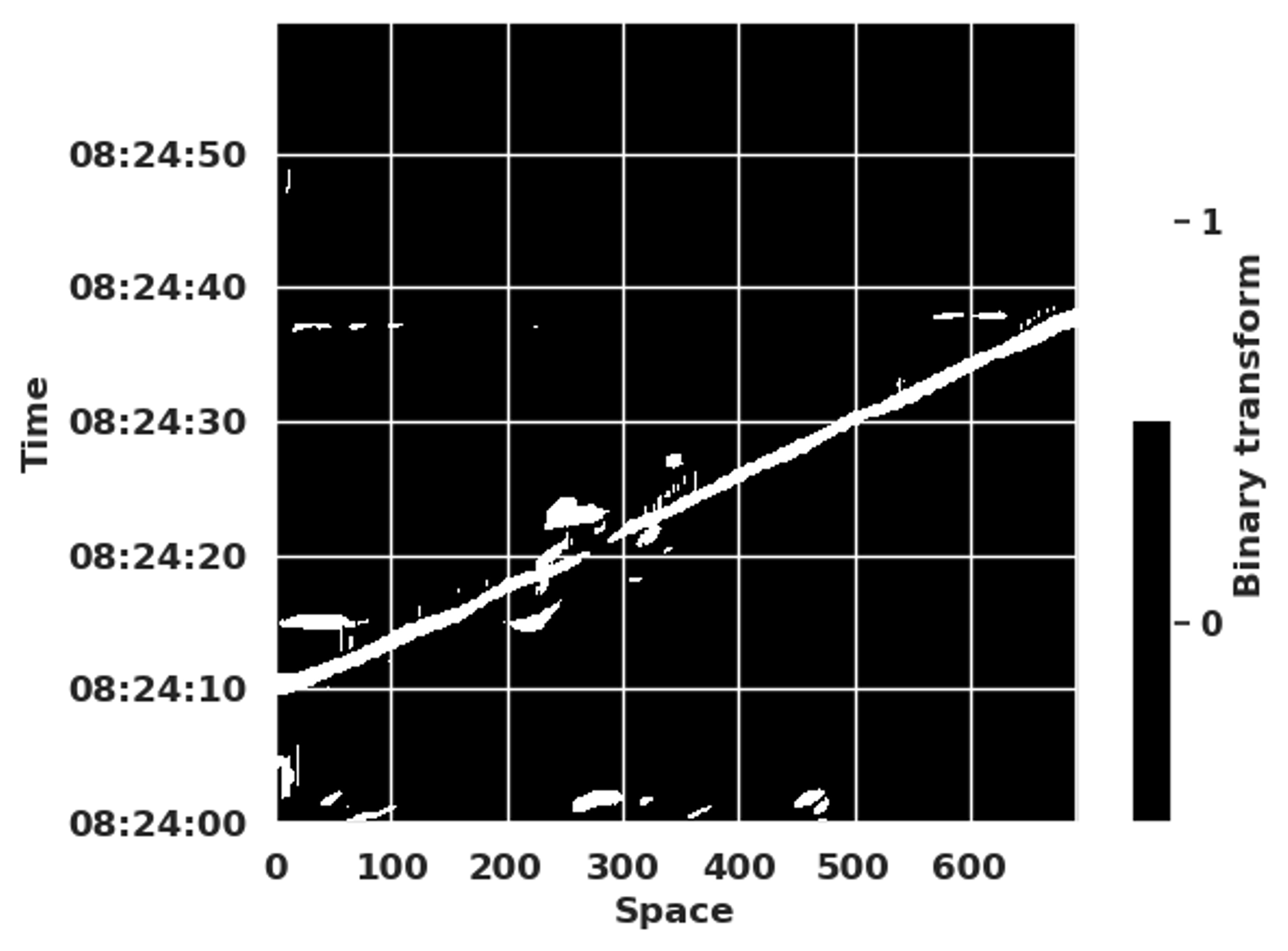}\label{fig:binary-low}}
    \hfill
    \subfloat[]{\includegraphics[trim={1.7in 0in 1.2in 0in}, clip, width=0.135\textwidth]{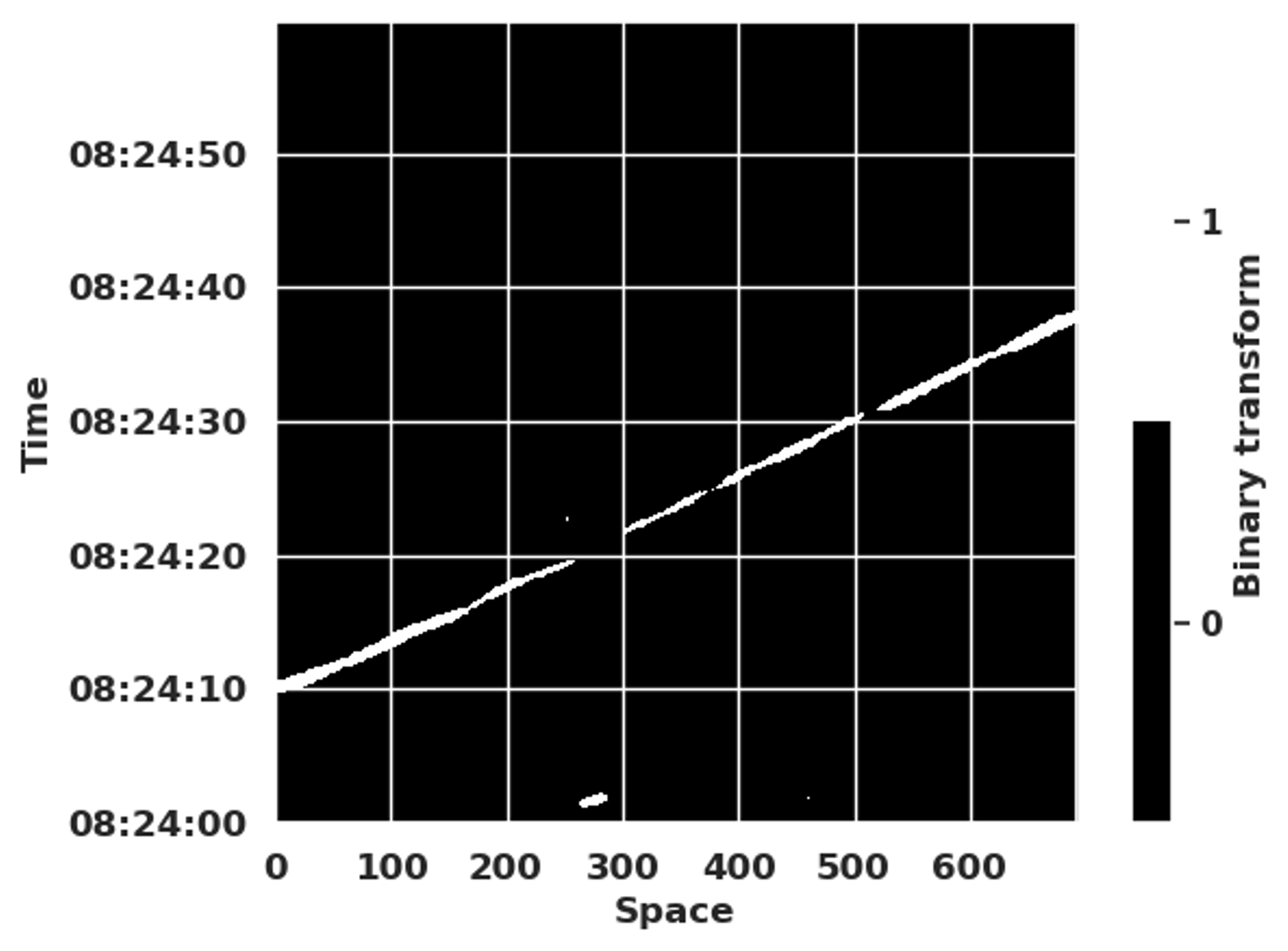}\label{fig:binary-medium}}
    \hfill
    \subfloat[]{\includegraphics[trim={1.7in 0in 0in 0in}, clip, width=0.165\textwidth]{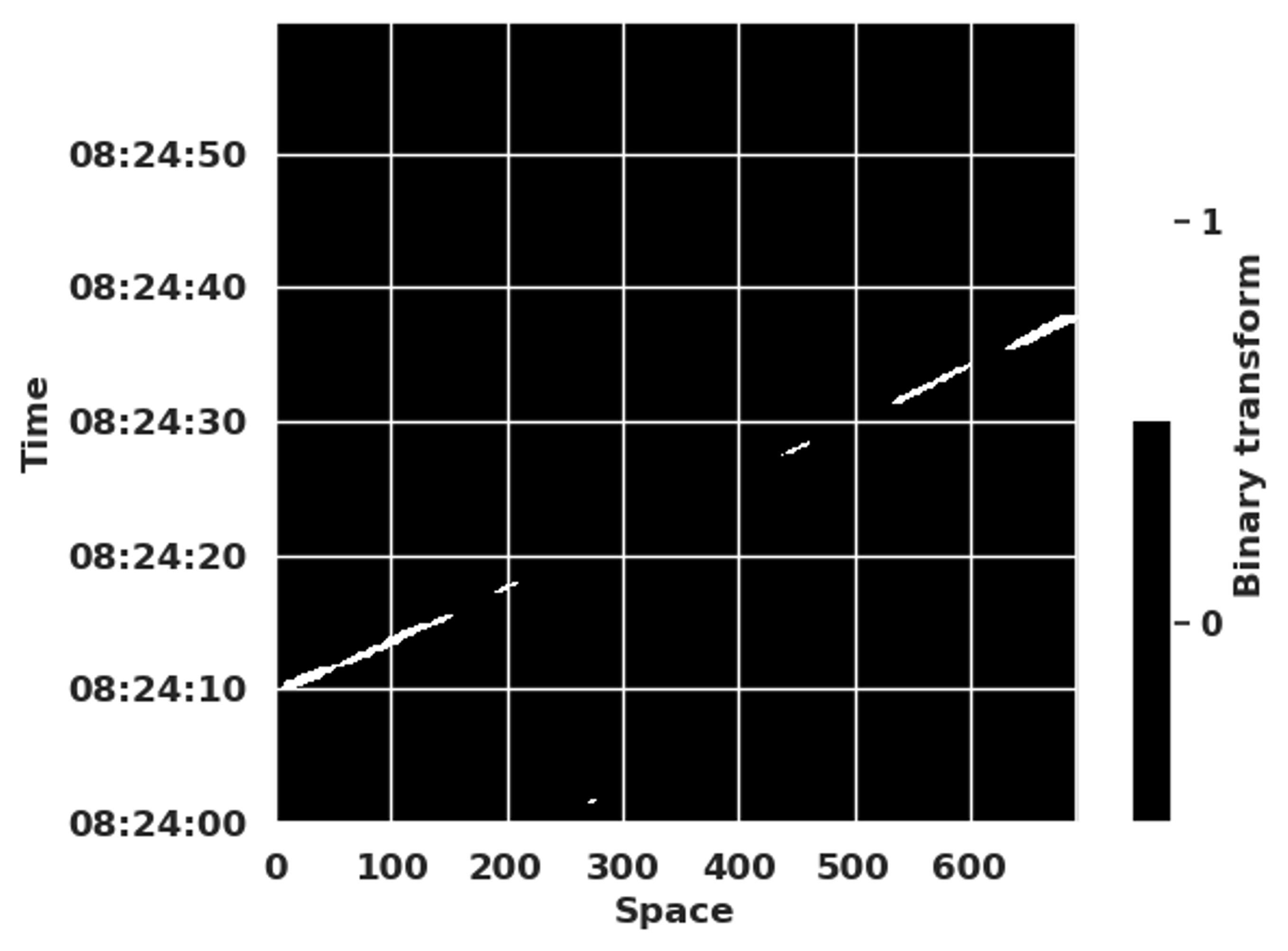}\label{fig:binary-high}}
    \caption{Binary transformation with different threshold values $\tau$. (a) Shows that $\tau=1 \times 10^{-8}$ is too low and retains much noise, whereas (c), $\tau=3 \times 10^{-8}$, is too high and the vehicle crossing signals is week. (b) $\tau=2 \times 10^{-8}$ is closer to the optimal value. The optimal threshold is fine tuned using a loss function together with an optimization algorithm.}
    \label{fig:binary}
\end{figure}

The image magnitude of the image is converted into a binary using a threshold value. Above the specified threshold, the pixel values are set to 1. Below the threshold they are set to 0. The threshold value $\tau$ is determined by analyzing the density distribution of the gradient magnitude image. Figure \ref{fig:binary} shows the binary images obtained with different threshold values. It must balance detection of vehicle crossings and avoiding noise. The optimal threshold is fine-tuned later using a loss function together with an optimization algorithm.

\subsection{Hough transform}

The Hough transform is a technique commonly used in image processing to detect straight lines \cite{duda1972use}. Considering our processed \gls{das} data as a binary image (Figure \ref{fig:binary}), the Hough transform works by converting points (binary 1-entry values) in the image space into a parameter space where lines can be identified more easily.

Considering an image, a straight line can be described by the Cartesian form $ y = mx + c $, where $ m $ is the slope and $ c $ is the $y$-intercept. However, this representation struggles with vertical lines, where $ m $ becomes infinite. Instead, we use polar form of the line:
\begin{equation}
    \rho = x \cos \theta + y \sin \theta. \label{eq:hough}
\end{equation}
Here, $ \rho $ is the perpendicular distance from the origin to the line, and $ \theta $ is the angle between the $x$-axis and the line's perpendicular. The origin in this context refers to the top-left corner of the spatiotemporal image. Every line in the image corresponds to a single point in the Hough space defined by $ (\rho, \theta) $. The range for $ \theta $ is typically from $-\frac{\pi}{2}$ to $\frac{\pi}{2}$ (or from 0 to $\pi$), while $ \rho $ ranges from $-\rho_{\text{max}}$ to $\rho_{\text{max}}$, where $ \rho_{\text{max}} $ is the diagonal length of the image. Then an array $ A(\rho, \theta) $ is created to accumulate votes for each possible line detected in the binary image. This array's dimensions are based on the resolution of $ \rho $ and $ \theta $. For each edge pixel in the image, the corresponding $ \rho $ is calculated for a range of $ \theta $ values, using equation (\ref{eq:hough}). The accumulator array  $ A(\rho, \theta) $  is updated by incrementing the value at the corresponding position. This process is repeated for every edge pixel in a so called voting process. High-value points in the accumulator array (called peaks) correspond to the most likely lines connecting the 1-entries in the binary image. Each peak corresponds to a line that has received multiple votes from different edge pixels, indicating that many pixels in the image are aligned. Finally, the parameters $ (\rho, \theta) $ can easily be converted back to Cartesian coordinates to draw the detected lines on the original image.

To improve computational efficiency, we use probabilistic Hough transform implemented in software library OpenCV \cite{matas2000robust, opencv2022opencv}. Probabilistic Hough transform is an optimized version of the original Hough transform. It randomly samples a subset of the edge pixels and only performs the voting process on this subset. This reduces the computational load while still detecting the most prominent lines in the image. In addition, it returns line segments directly in the image space rather than just returning $(\rho, \theta)$ values for infinite lines.

The probabilistic Hough transform has five parameters: (1) the resolution of $ \rho $, (2) the resolution of $ \theta $, (3) the minimum number of votes required to recognize a line during the voting process, (4) the minimum line length, and (5) the maximum gap between points to be considered part of the same line. We set $\rho$-resolution to 1, and $\theta$-resolution to achieve a velocity accuracy of 0.1 km/h at 85 km/h. Parameters (3), (4), and (5) are configured so that detected lines cover at least 70\% of the bridge length, with a maximum point gap of 20\%.

\begin{figure}[htbp]
    \centering
    \includegraphics[width=0.35\textwidth, trim={0in, 0in 1.2in 0in}, clip]{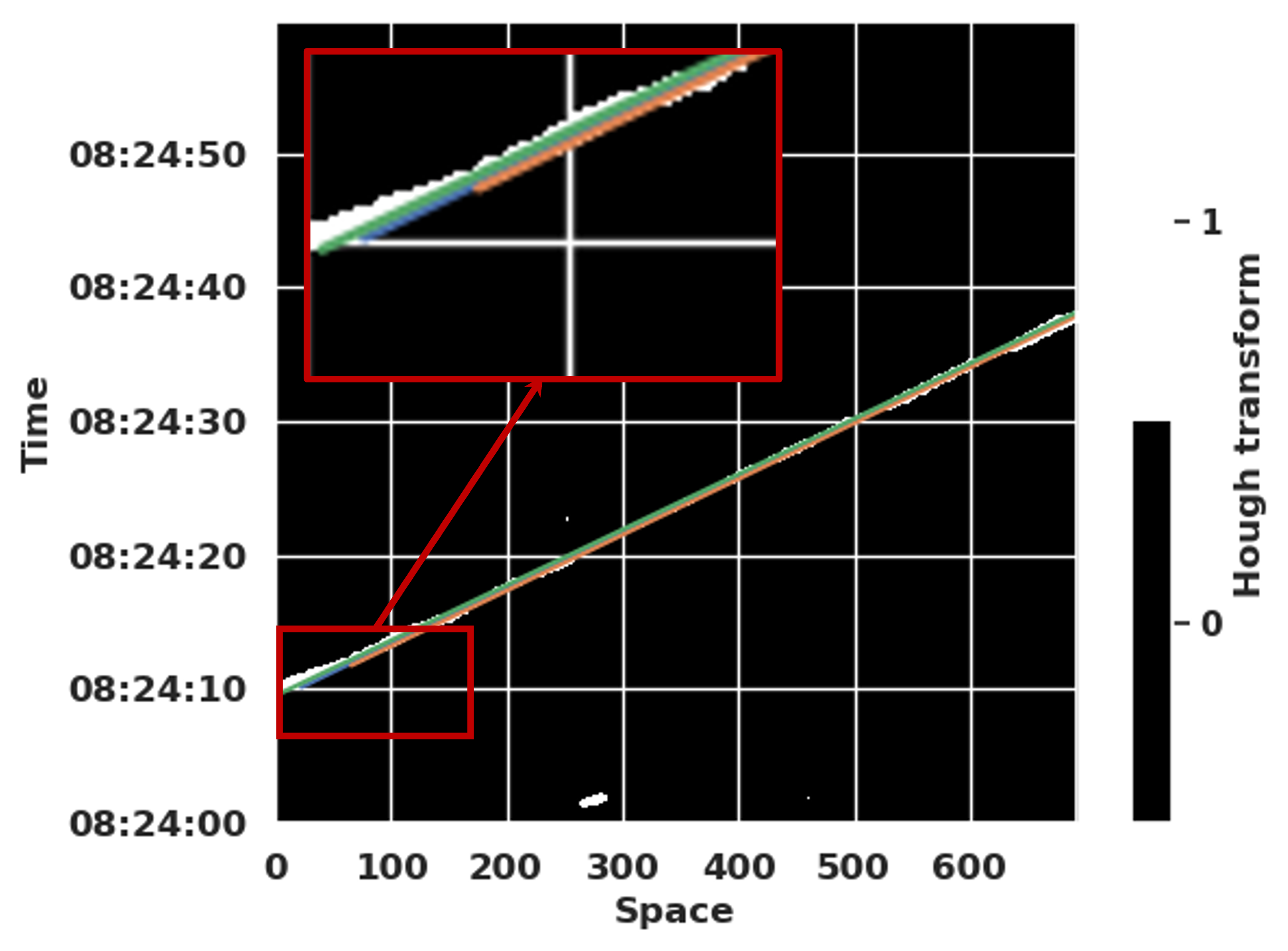}
    \caption{Result from Hough transform with three detected line segments, representing the same vehicle. The associated speeds are 86.0, 86.3 and 86.8 km/h.}
    \label{fig:hough}
\end{figure}

Figure \ref{fig:hough} shows three detected line segments from the Hough transform for one car crossing the bridge in our \gls{das} data. We need to consolidate these line segments into a single segment to represent the vehicle crossing. This is done next by line clustering.

\subsection{\gls{dbscan}}

\begin{figure}[htbp]
    \centering
    \includegraphics[width=0.35\textwidth]{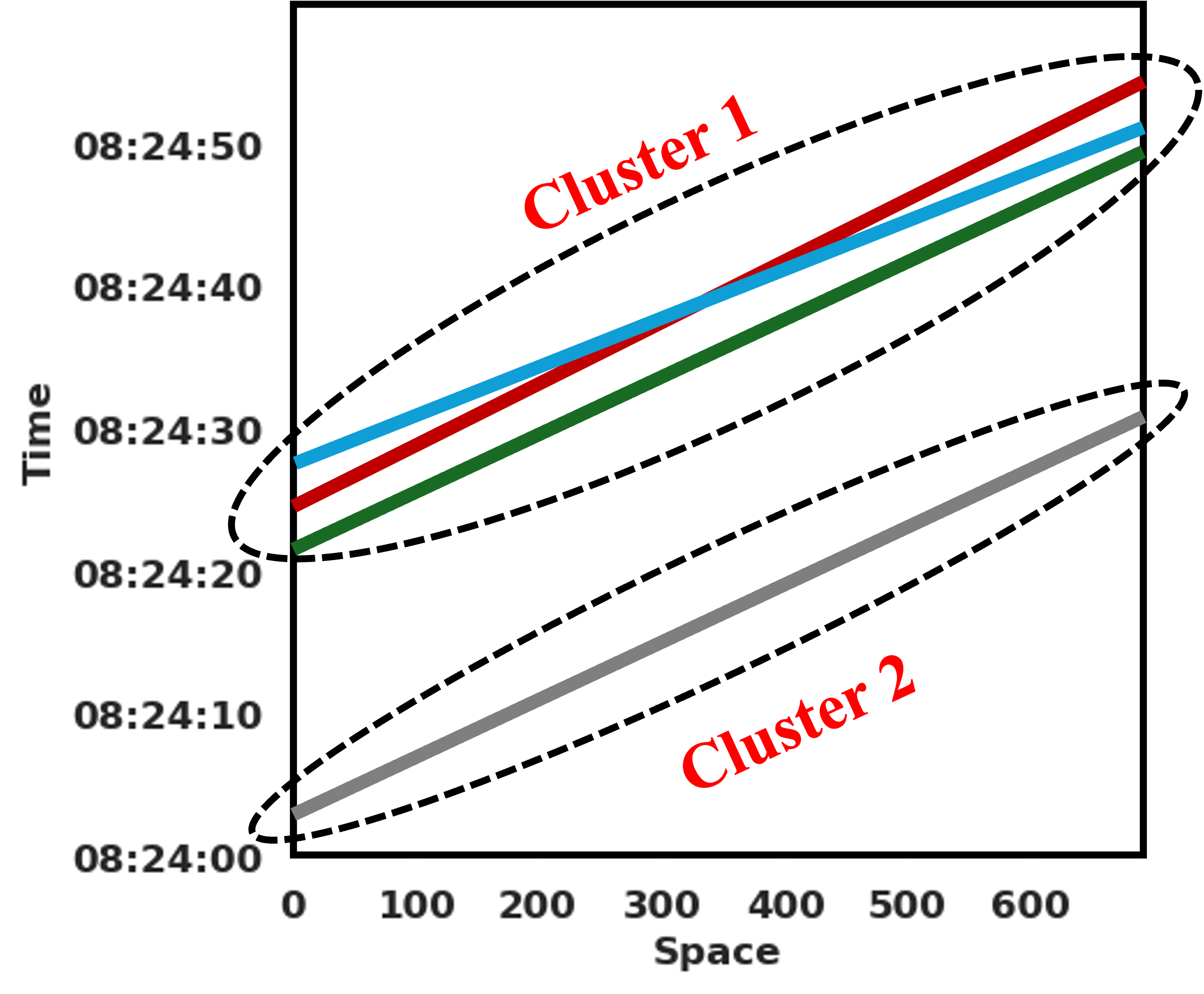}
    \caption{Example of line segments clustering using \gls{dbscan}. Line segments that are within a distance of $\epsilon$ from each other are considered neighbors and grouped to a cluster.}
    \label{fig:dbscan}
\end{figure}

The Hough transform may identify multiple line segments for one single vehicle crossing in the \gls{das} dataset, as shown in Figure \ref{fig:hough}. These segments are typically very close to one another. We first extrapolate the detected line segments to the full length of the bridge, then consolidate these into a single segment per vehicle using \gls{dbscan} \cite{ester1996density} with a customized distance metric. Following is a brief description of the \gls{dbscan} algorithm for our case.

Given a set of line segments resulting from the Hough transform, \gls{dbscan} algorithm effectively clusters close line segments. It relies on two main parameters: $\epsilon$ which defines the maximum distance between two segments to be considered as neighbors, and $\textit{minSegs}$ which is the minimum number of line segments required to form a cluster. \gls{dbscan} operates as follows:

\begin{enumerate}
    \item Core line segments: A line segment $ l $ is classified as a core line segment if there are at least $\textit{minSegs}$ line segments (including $ l $) within a distance $\epsilon$ from $ l $. The $\epsilon$-neighborhood of $ l $ is denoted $ N(l) $. For $ l $ to be a core line segment, the number of line segments in $ N(l) $ (size of $N(l)$: $|N(l)|$) must satisfy $ |N(l)| \geq \textit{minSegs}$.
    \item Border line segments: A line segment $ l' $ is considered a border line segment if it is within the $\epsilon$ neighborhood of a core line segment $ l $, but $ l' $ itself does not have enough line segments in its neighborhood to be a core line segment. Formally, $ l' \in N(l) $ but $ |N(l')| < \textit{minSegs}$.
    \item Noise line segments: line segments that are neither core line segments nor border line segments are classified as noise. These line segments do not belong to any cluster and are considered outliers.
\end{enumerate}

The algorithm starts by selecting an arbitrary line segment and checking whether it is a core line segment by examining its $\epsilon$-neighborhood. If it is a core line segment, a new cluster is started. All line segments within its $\epsilon$-neighborhood are added to this cluster, and their neighbors are also explored recursively to find additional line segments belonging to the cluster. This process continues until all line segments in the cluster are discovered. Next, the algorithm moves to the next un-visited line segment and repeats the process, identifying new clusters as it proceeds. The process continues until all line segments in the dataset have been visited.

In our study, we have not observed cases where a single line segment from the Hough transform corresponds to one vehicle. Typically, due to noise, multiple line segments represent a single vehicle. However, in ideal conditions with high quality data, it is theoretically possible to detect only one line segment per vehicle. Therefore, we set $\textit{minSegs} = 1$, allowing for the possibility of a single line segment being detected for a vehicle crossing. Consequently, all line segments are treated as core points, and \gls{dbscan} here behaves similarly to single-linkage hierarchical clustering or nearest neighbor clustering.

The remaining challenge is to define a distance metric between two line segments to determine if they are within $\epsilon$ of each other. The distance between two line segments can be defined as the average area between them over the range of the bridge. Mathematically, the distance between line segments $l_1$ and $l_2$ over the spatial indices $s$ of the bridge is:

\begin{equation}
    D(l_1,l_2) = \frac{1}{693} \int_{0}^{693} |l_1(s) - l_2(s)| ds.
    \label{eq:distance}
\end{equation}

The optimal $\epsilon$ is determined by fine-tuning using a loss function and an optimization algorithm, which is described in the next section.

\subsection{Loss function and parameters tuning}

\subsubsection{Loss function}

The expected output of the proposed method is a set of line segments described by 2-entry vector timestamps of vehicles crossing two ends of the bridge. The accuracy of the output is evaluated by comparing it with the ground truth line segments made in-situ with a camera. Define sets $\mathcal{P}=\{p_1, p_2, ..., p_n\}$ and $\mathcal{Q}=\{q_1, q_2, ..., q_m\}$ for the ground truth and the predicted line segments of vehicles crossing the bridge, respectively. We define neighborhood sets of every ground truth $p_i$, called $S_{p_i}$, as follows:
\begin{equation}
    S_{p_i} = \{ q_j \in \mathcal{Q} |  d(p_i, q_j) \le d(p_k, q_j) \forall p_k \in \mathcal{P}, p_k \neq p_i\},
\end{equation}
where $d(p_i, q_j)$ is the distance between $p_i$ and $q_j$. In case of equality, we randomly assign to one of the sets. When $S_{p_i}$ is empty, i.e. $|S_{p_i}|=0$, no prediction is associated with $p_i$, resulting in a \gls{fn}. When $|S_{p_i}| = 1$, there is only one prediction for $p_i$, which is considered a \gls{tp}. When $|S_{p_i}| > 1$, there are multiple predictions for $p_i$, resulting in one \gls{tp} (the one that is closest to $p_i$) and $|S_{x_i}| - 1$ \gls{fp}. Based on this, we use a loss function:
\begin{align}
    f(\mathcal{P}, \mathcal{Q}) = & \frac{1}{n}\sum_{i=1}^{n} \Bigg[ \mathbf{1}_{|S_{p_i}| = 0} \cdot \alpha                                                              \\
    +         & \mathbf{1}_{|S_{p_i}| > 0} \cdot \left(\min_{q_i \in S_{p_i}} \{d(p_i, q_i)\} + \beta \cdot (|S_{p_i}| - 1)\right) \Bigg], \nonumber
\end{align}
where $\alpha$ is the \gls{fn} penalty and $\beta$ is the \gls{fp} penalty. To demonstrate how the loss function works, considering an illustrative case where $p_i$ and $q_i$ are simple scalars. Let $\mathcal{P} = \{10, 20, 30\}$,  $\mathcal{Q} = \{8, 19, 22, 23\}$, we infer $S_{10} = \{8\}$, $S_{20} = \{19, 22, 23\}$, and $S_{30} = \emptyset$. With $\alpha=3$, $\beta=3$ and $d(p_i, q_j)=|p_i - q_j|$, the loss:
\begin{align*}
    f(\mathcal{P}, \mathcal{Q})
     & = \frac{1}{3} \left[|8-10| + (|19-20| + 2 \cdot \beta) + \alpha \right]     \\
     & = \frac{1}{3} \left[2 + (1 + 2 \cdot 3) + 3 \right]=\frac{1}{3} \cdot 12 = 4.
\end{align*}
Given that $p_i$ and $q_i$ are line segments in our actual study, we use the distance measure defined in equation~\eqref{eq:distance} to calculate $d(p_i, q_i)$. The loss function hence reflects the average time difference, in seconds, between the ground truth and the predicted lines. The lower the loss, the better the performance of the method. The loss function is used to tune the parameters of the method to achieve the best performance. We set $\alpha=5$ and $\beta=5$, i.e. every \gls{fn} and \gls{fp} is penalized by 5 seconds.

\subsubsection{Parameters tuning}

Manual parameter tuning is labor-intensive. It is therefore helpful to have a systematic way of choosing the parameters of the algorithm. Based on a loss function, we use the \gls{tpe} algorithm to tune select parameters. The \gls{tpe} algorithm is a Bayesian optimization algorithm that uses a probabilistic model to predict the loss function based on the previous evaluations \cite{bergstra2011algorithms}. The following is a brief overview of the \gls{tpe} algorithm.

Let \( w \) represent the parameter vector, and \( f(\mathcal{P}, \mathcal{Q}) = f(w) \) is the loss function that we aim to minimize over the parameter space \( W \). We first split $W$ into two sets \( \mathcal{L} \) and \( \mathcal{G} \). They are parameters corresponding to good performance (i.e., loss values below a performance quantile \( f^* \)) and poor performance (i.e., loss values above  \( f^* \)).
\begin{equation}
    \mathcal{L} = \{ w \in W : f(w) \leq f^* \},
    \hspace{4mm}
    \mathcal{G} = \{ w \in W : f(w) > f^* \}.
\end{equation}
Here, this quantile \( f^* \) is set to a 15\% fraction among the lowest values. \gls{tpe} then estimates two probability density functions for the parameters conditioned on whether they resulted in good or poor performance:
\begin{equation}
    l(w) = \textit{pdf}(w | f(w) \leq f^*),
    \hspace{8mm}
    r(w) = \textit{pdf}(w | f(w) > f^*).
\end{equation}

These densities \( l(w) \) and \( g(w) \) are modeled using non-parametric \gls{kde}, which approximates the distribution of the good and poor sets of parameters by placing Gaussian kernels on top of each evaluated point. Next, the \gls{tpe} algorithm chooses new parameters \( w_{\text{new}} \) by maximizing the ratio $\frac{l(w)}{g(w)}$:
\begin{equation}
    w_{\text{new}} = \arg \max_w \frac{l(w)}{g(w)}.
\end{equation}

The ratio \( \frac{l(w)}{g(w)} \) represents how much more likely it is that a given parameter configuration \( w \) belongs to the set of good configurations \( \mathcal{L} \) compared to the set of poor configurations \( \mathcal{G} \). By maximizing this ratio, the algorithm favors parameter configurations that are more likely to result in good loss values. 
Once the ratio \( \frac{l(w)}{g(w)} \) is maximized, the corresponding parameter configuration \( w_{\text{new}} \) is evaluated by the loss function \( f(w_{\text{new}}) \), and the results are added to the dataset of past evaluations. The algorithm then updates its density estimations \( l(w) \) and \( g(w) \) and repeats this process until a termination criterion (e.g., a time budget or a maximum number of trials) is met.

We use the \gls{tpe} algorithm to tune parameters of the proposed method for which we lack prior knowledge to determine. The parameters include the cutoff frequency $f_c$ in \gls{lpf}, the threshold $\tau$ for binary transformation and neighborhood radius $\epsilon$ in \gls{dbscan}. The tuning is done using the Optuna library \cite{akiba2019optuna}. The tuning showed that the loss was most sensitive to the threshold $\tau$ in binary transformation.

\subsection{Real-time deployment}

\begin{figure}[htbp]
    \centering
    \includegraphics[width=3in]{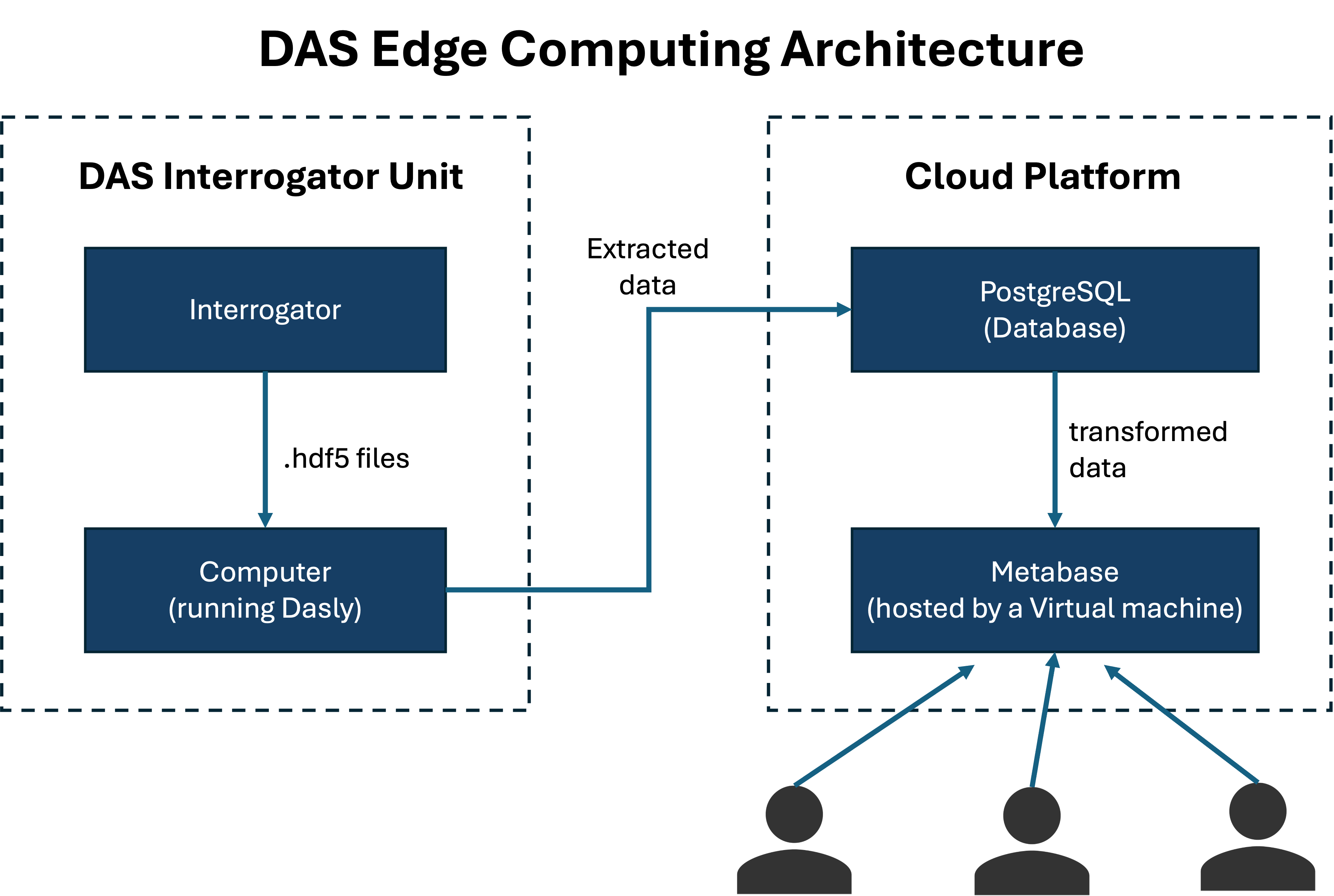}
    \caption{Real-time deployment architecture. The interrogator generates  \gls{hdf5} files every 10 seconds. The data is processed in the edge computer by Dasly. The output is sent to a cloud platform for storing and visualization.}
    \label{fig:edge-compute-architecture}
\end{figure}

We have developed and deployed an edge-computing workflow for the vehicle detection and attribute estimation on the bridge. The core elements of the workflow are summarized in Figure \ref{fig:edge-compute-architecture}. The development and testing phases were conducted on a local workstation, while the final algorithm was deployed on the edge computer, which is connected to the \gls{iu} at the bridge. Both machines are capable of processing \gls{das} data. The workstation has 12 cores of CPU, compared to 8 cores on the edge computer. The workstation also has more RAM and a dedicated GPU. On the other hand, the edge computer has a higher clock speed (3.6 GHz base and 5.0 GHz boost compared with 3.5 GHz base and 4.4 GHz boost).

A Python package named  Dasly is developed based on the proposed methodology and is hosted at \url{github.com/truongphanduykhanh/dasly}. The \gls{iu} generates new \gls{hdf5} files every 10 seconds, each contains the latest \gls{das} data. A file system monitoring tool, Watchdog \cite{schoentgen2023watchdog}, triggers Dasly whenever a new file is created.  Dasly loads the data, runs the preprocessing, Hough transform, and clustering with \gls{dbscan}. The output, including line coordinates and derived metrics like speed and location, is sent to a PostgreSQL database hosted in a cloud computing platform. Visualization is handled by Metabase, which connects to the database to provide a real-time dashboard displaying vehicle counts, average speeds, and system latency.

The primary consideration for edge deployment is that the algorithm must process data fast enough to handle the continuous written \gls{hdf5} files. There are two key parameters influence deployment: \textit{batch\_gap} and \textit{batch\_duration}. \textit{batch\_gap} is the frequency at which Dasly runs. Setting the \textit{batch\_gap} to 10 seconds means Dasly is triggered with every new \gls{hdf5} file. While setting the \textit{batch\_gap} to 60 seconds means Dasly is triggered with every 6th new \gls{hdf5} file. The higher the \textit{batch\_gap} is, the higher latency the system has. On the other hand, the second parameter \textit{batch\_duration} is the length of data processed per run. Setting the \textit{batch\_duration} to 10 seconds means that Dasly processes the last 10 seconds of data. While setting the \textit{batch\_duration} to 60 seconds means Dasly processes the last 60 seconds of data. The higher the \textit{batch\_duration} is, the more data is covered at one run but the longer the algorithm will run. The trade off between the two parameters is important to consider when deploying the system.

It is crucial to ensure that \textit{batch\_duration} is at least equal as long as \textit{batch\_gap} to avoid data loss. If \textit{batch\_duration} exceeds \textit{batch\_gap}, the system processes overlapping data from consecutive batches, which helps to reinforce detection accuracy. Each line detected in a batch is assigned a unique ID. If one line in the batch is close enough to one of the lines in the adjacent batch (Equation~\eqref{eq:distance} and Figure~\ref{fig:dbscan}), it will be assigned the same ID. Ideally, \textit{batch\_gap} should be minimized to reduce latency, while \textit{batch\_duration} should be maximized to enhance robustness, as long as processing time remains within the \textit{batch\_gap} limit.

Dasly’s execution time in the workstation increases approximately linearly with \textit{batch\_duration}, taking about 6 seconds to process a 60-second batch. This is well within the 10-second limit, leaving a margin for potential temporary overloads. Therefore, \textit{batch\_duration} is set to 60 seconds, with a \textit{batch\_gap} of 10 seconds, resulting in a 50-second overlap between consecutive batches.

\section{Result and discussion}
\label{sec:results}

\begin{figure}[htbp]
    \centering
    \includegraphics[width=3in]{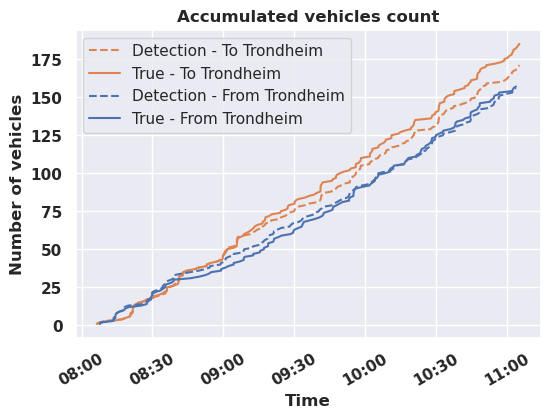}
    \caption{Predicted and true lines in 3 hours of data.}
    \label{fig:predict-true-lines}
\end{figure}

A total of 342 vehicles were recorded crossing the bridge, with 185 traveling to Trondheim and 157 in the opposite direction. Figure \ref{fig:predict-true-lines} demonstrates that the predicted lines from our workflow closely match the true lines observed in the camera footage, indicating the effectiveness of the proposed vehicle detection method. In the Trondheim direction (orange line), a widening gap between the predicted and true lines suggests the algorithm missed a few vehicles over time. Closer analysis reveals that these \gls{fn}s often occur when cars follow closely behind large trucks. The strong \gls{das} signal generated by trucks masks the smaller vehicles behind, making it hard to detect them. In the from Trondheim direction (blue line), the algorithm detects more lines than the actual vehicles between 08:30 and 09:00. This over-detection is attributed to large trucks generating significant noise, resulting in multiple \gls{fp}s.

\begin{figure}[htbp]
    \centering
    \subfloat[]{\includegraphics[trim={0in 0in 0in 0in}, clip, width=0.7\linewidth]{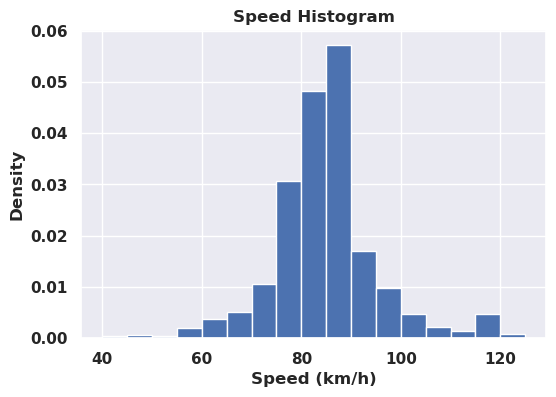}\label{fig:speed-hist}}
    \hfill
    \subfloat[]{\includegraphics[trim={0in 0in 0in 0in}, clip, width=0.52\linewidth]{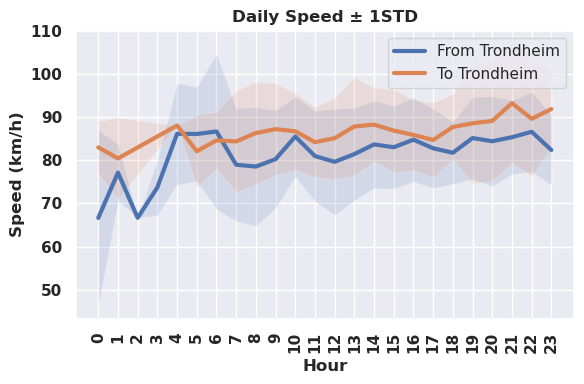}\label{fig:speed-daily}}
    \hfill
    \subfloat[]{\includegraphics[trim={0in 0in 0in 0in}, clip, width=0.48\linewidth]{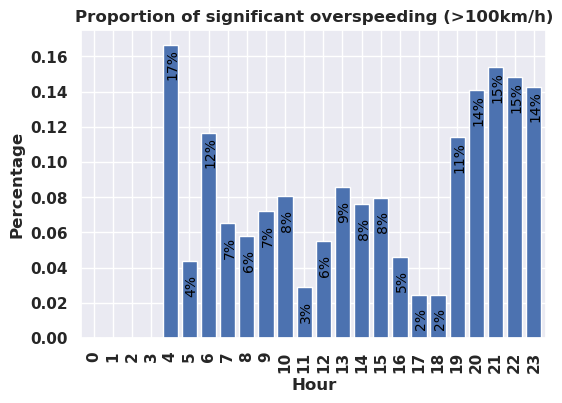}\label{fig:speed-overspeed}}  
    \caption{Speed statistics. (a) Histogram of the vehicle speeds. The most common range of the speed is between 75 and 90 km/h. (b) Hourly average speeds over 24-hour in one day. The shaded region is $ \pm 1$ standard deviation from the average. The speed to Trondheim (downhill) is higher than the one from Trondheim (uphill). (c) Proportion of vehicles driving over the speed of 100 km/h. Most happen at evening and early morning.}
    \label{fig:speed}
\end{figure}

Figure \ref{fig:speed} presents velocity statistics for the detected vehicles. The histogram (Figure \ref{fig:speed-hist}) indicates that most vehicles travel between 75 and 90 km/h, with a small peak near 115-120~km/h. In Norway, driving licenses are confiscated for speeds of 116~km/h or more in an 80~km/h zone, with a 3~km/h tolerance applied. This suggests that some drivers are aware of the threshold and aim to drive just below 120 km/h. These drivers might be willing to pay the fine for the speeding but not to lose their license \cite{jogensen2002car}.

The hourly average speeds over a 24-hour period (Figure \ref{fig:speed-daily}) show that vehicles heading toward Trondheim generally travel faster than those coming from Trondheim, likely due to the bridge’s downward slope in that direction. Additionally, the proportion of vehicles exceeding 100~km/h (Figure \ref{fig:speed-overspeed}) appears to peak in the evening (19:00 -- 23:00) and early morning (04:00), when traffic is lighter.

\begin{figure}[htbp]
    \centering
    \includegraphics[width=3in]{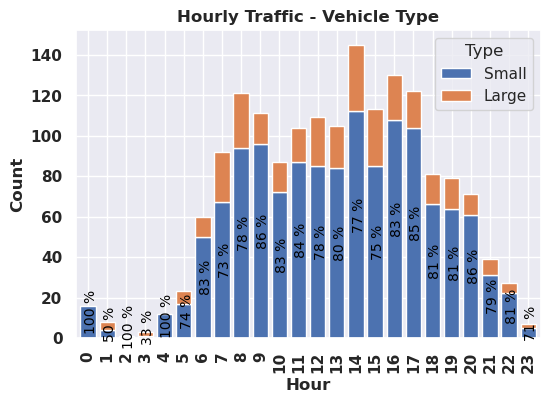}
    \caption{Traffic count by vehicle type. The number of cars is significantly higher than the number of trucks.}
    \label{fig:traffic-count-type}
\end{figure}

It is interesting to split the traffic in light and heavy vehicles. In a simple analysis we computed the average signal strength and frequency content for the detected lines (Figure \ref{fig:fft}). This was done for the training data with both camera and \gls{das} data, and by doing so we know if a vehicle is a truck or a car. Signal characteristics are computed for \gls{das} data along the detected lines using our algorithm.  Figure \ref{fig:traffic-count-type} shows the traffic count by vehicle type. The number of small vehicles (passenger cars, vans) is accounting for around 80\% of the traffic. This is consistent with the fact that the bridge is primarily used by commuters and tourists.

\begin{figure}[htbp]
    \centering
    \subfloat[]{\includegraphics[trim={0in 0in 0in 0in}, clip, width=0.465\linewidth]{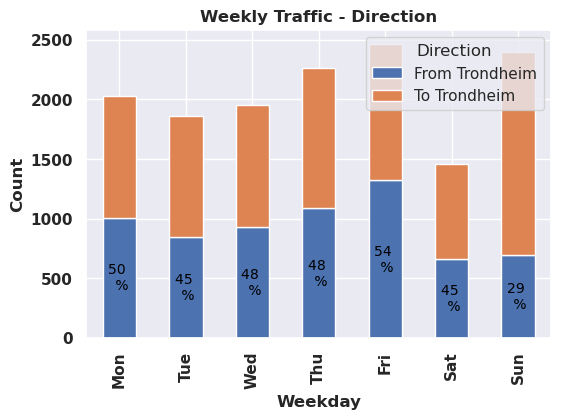}\label{fig:weekly-count}}
    \hfill
    \subfloat[]{\includegraphics[trim={0in 0in 0in 0in}, clip, width=0.53\linewidth]{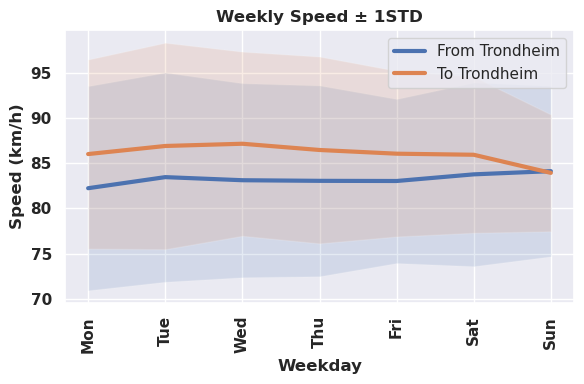}\label{fig:weekly-speed}}
    \caption{Weekly traffic summary. (a) Number of vehicles crossing the bridge every day of the week. (b) Speed of the vehicles crossing the bridge every day of the week. The shaded region is $ \pm 1$ standard deviation from the average.}
    \label{fig:week}
\end{figure}

During weeks of testing, we observed 1,500 to 2,500 vehicles per day, with noticeable variations in weekend traffic (Figure \ref{fig:week}). Specifically, there is a higher volume of cars leaving Trondheim on Fridays and returning on Sundays, which is expected as the bridge leads to an area of cabins and plenty of leisure activities. Regarding speed, vehicles traveling toward Trondheim drive faster than those heading in the opposite direction (consistent with the slope of the bridge), but there is no significant difference in speed between weekdays and weekends.

\begin{figure}[htbp]
    \centering
    \subfloat[]{\includegraphics[trim={0in 0in 1.1in 0in}, clip, width=0.52\linewidth]{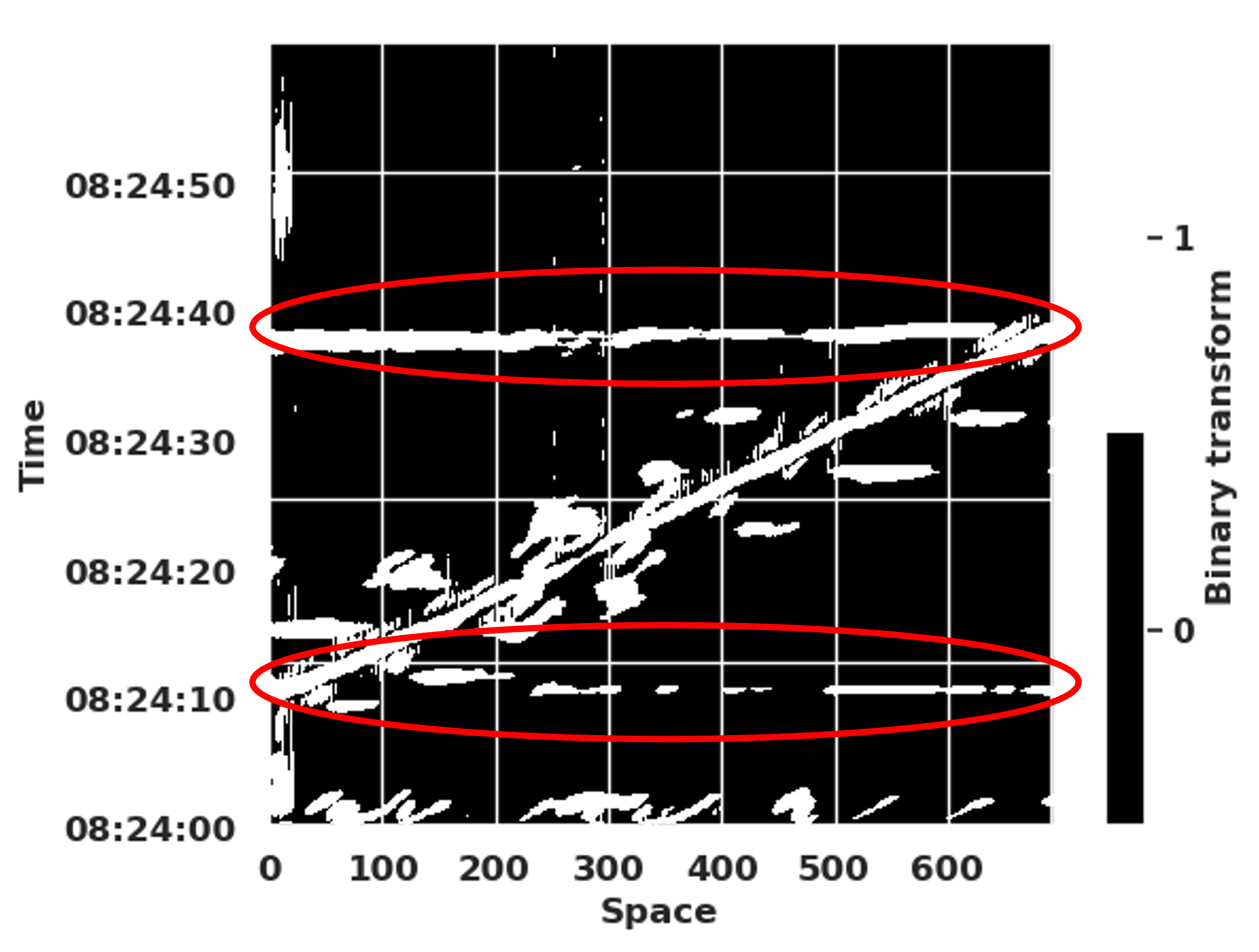}\label{fig:s-wave-binary}}
    \hfill
    \subfloat[]{\includegraphics[trim={1.7in 0in 0in 0in}, clip, width=0.48\linewidth]{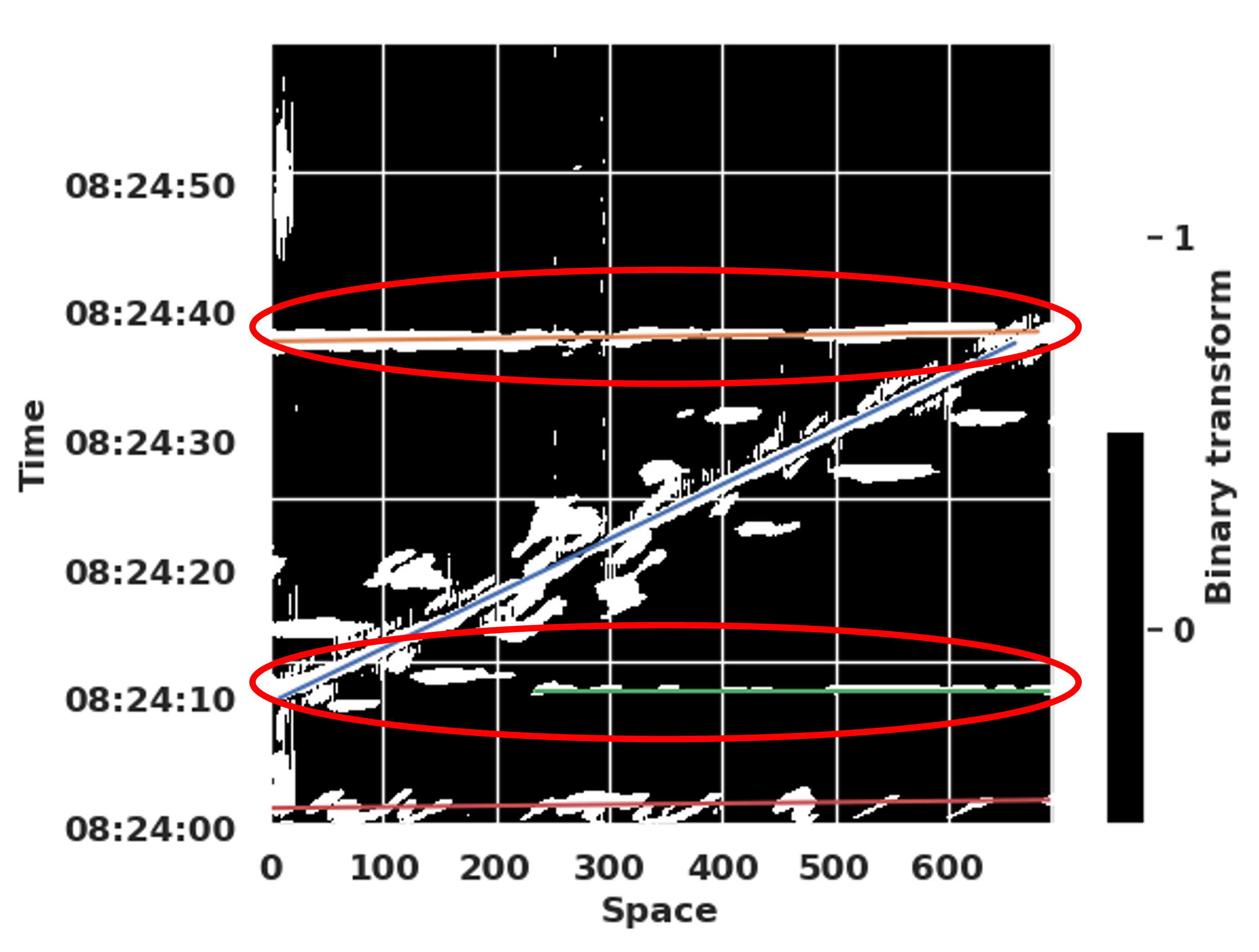}\label{fig:s-wave-hough}}
    \caption{S-wave detection. (a) Binary transformation of the \gls{das} data. (b) Detected S-waves using Hough transform and \gls{dbscan}.}
    \label{fig:s-wave}
\end{figure}

Apart from detecting the vehicles, our proposed method can also conveniently detect the S-waves in the concrete, triggered by the vehicle entering the bridge (Figure \ref{fig:s-wave}). We clearly observe a speed of $\simeq$3000~m/s for the S-wave, consistent with known values for concrete. Tracking parameters such as the S-wave speed over time - can be beneficial for bridge structural health monitoring.

\section{Conclusion}
\label{sec:conclusion}

In this study, we presented a novel approach for real-time vehicle detection and velocity estimation using \gls{das} data, focusing on the Åstfjord bridge in Norway. Our method integrates the Hough transform for line detection and \gls{dbscan} for clustering, enabling accurate detection of vehicles in both directions across the bridge. By comparing \gls{das} data with camera footage, we demonstrated that our system is highly effective in identifying vehicle crossings and estimating their speeds, even in challenging scenarios where many vehicles run close to each other.

The analysis of traffic patterns over a 24-hour period and across different days of the week revealed consistent trends, including higher speeds towards Trondheim and increased weekend traffic. We also observed that vehicles tend to drive faster in the evening and early morning when traffic is lighter. Moreover, we successfully classified vehicles into light and heavy categories based on \gls{das} signal strength and frequency content, providing insights into traffic composition.

A significant contribution of this work is the edge computing deployment, which processes \gls{das} data in real-time and streams results to a cloud based visualization platform. This efficient system is capable of handling large volumes of data while maintaining low latency, ensuring timely analysis for traffic monitoring.

Additionally, the methodology shows promise beyond traffic applications, as it detected structural responses such as shear waves in the bridge’s foundation. This highlights its potential for applications in structural health monitoring and other domains like seismic or wildlife tracking, where similar signal characteristics are present. Future work could further improve the classification of vehicles, particularly in challenging scenarios where cars follow trucks, and extend the methodology to long-distance roads, where vehicles may travel at more variable speeds.

\section*{Acknowledgments}

This work is supported by the Norwegian Research Council SFI Centre for Geophysical Forecasting grant no. 309960. This project has also received funding from the European Union’s Horizon 2020 research and innovation program under the Marie Skłodowska-Curie grant agreement no. 101034240. We want to acknowledge Trondheim fylkeskommune for giving us access to the bridge and \gls{cgf} team for data acquisition. To improve readability and quality of language, this writing has been grammatically revised using ChatGPT-4o.

\bibliography{references/references}
\bibliographystyle{IEEEtran}

\newpage
\section{Biography}

\begin{IEEEbiography}[{\includegraphics[width=1in,height=1.25in,clip,keepaspectratio]{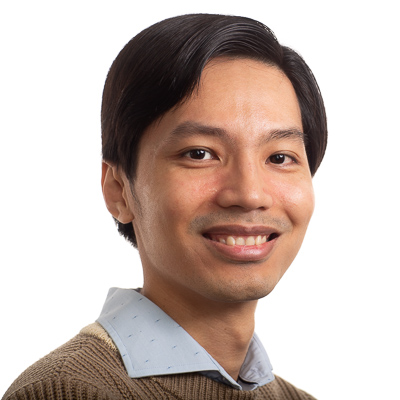}}]{Khanh Truong} is currently working as a PhD candidate at Mathematical Science department at \acrlong{ntnu}. He works on applying statistics and machine learning techniques to extract desired knowledge from \gls{das} data.\end{IEEEbiography}

\begin{IEEEbiography}[{\includegraphics[width=1in,height=1.25in,clip,keepaspectratio]{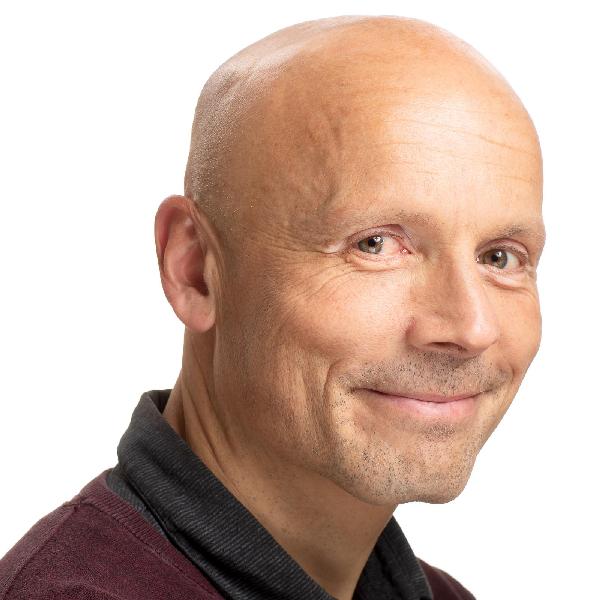}}]{Jo Eidsvik} is a Professor of Statistics at the Department of Mathematical Sciences, \acrlong{ntnu}. He works on spatial and computational statistics.
\end{IEEEbiography}

\begin{IEEEbiography}[{\includegraphics[width=1in,height=1.25in,clip,keepaspectratio]{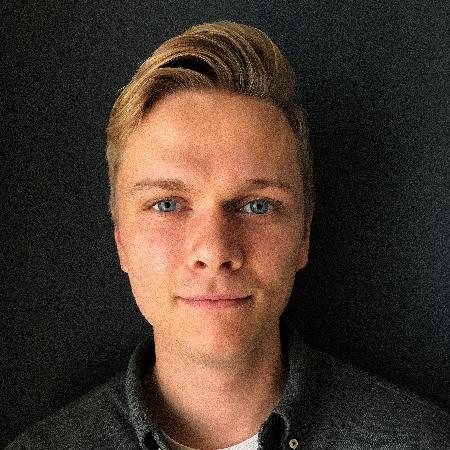}}]{Robin Andre Rørstadbotnen} is a Post-Doc Fellow at the Department of Electronic Systems, Norwegian University of Science and Technology. He works on signal processing for various DAS applications. 
\end{IEEEbiography}

\vfill

\end{document}